\documentclass[onecolumn,final]{elsart3p}
\usepackage{epsfig}

\usepackage{amssymb}
\usepackage{amsfonts}
\usepackage{dcolumn}
\journal{Nuclear Physics A}

\begin{document}

\begin{frontmatter}



\title{The role of the Pauli principle in three-cluster systems composed of identical clusters\thanksref{support}}
 \thanks[support]{This work was partly supported by the Program of Fundamental research of the Physics and Astronomy Department of
the National Academy of Sciences of Ukraine.}


\author{Yu. A. Lashko},
\ead{lashko@univ.kiev.ua}
\author{G. F. Filippov}
\address{Bogolyubov Institute for Theoretical Physics, 14-b Metrolohichna str., 03680, Kiev, Ukraine}

\begin{abstract}
Within the microscopic model based on the algebraic version of the
resonating group method the role of the Pauli principle in the
formation of continuum wave function of nuclear systems composed
of three identical $s$-clusters has been investigated. Emphasis is
placed upon the study of the exchange effects contained in the
genuine three-cluster norm kernel. Three-fermion, three-boson,
three-dineutron ($3d'$) and $3\alpha$ systems are considered in
detail. Simple analytical method of constructing the norm kernel
for $3\alpha$ system is suggested. The Pauli-allowed basis
functions for the $3\alpha$ and $3d'$ systems are given in an
explicit form and asymptotic behavior of these functions is
established. Complete classification of the eigenfunctions and the
eigenvalues of the $^{12}$C norm kernel by the
$^8$Be$=\alpha+\alpha$ eigenvalues has been given for the first
time. Spectrum of the $^{12}$C norm kernel is compared to that of
the $^{5}$H system.
\end{abstract}

\end{frontmatter}

\section{Introduction}
\label{intro}

The question of the role of the Pauli principle in three-cluster
systems goes back a long way in history. The $^{12}$C nucleus has
been established to exhibit three-alpha-cluster structure forty
years ago. Since then many microscopic, macroscopic and
semi-microscopic cluster models have been applied to analyze the
structure of the ground and excited states of this nucleus. In
particular, microscopic $3\alpha$ calculation was performed within
the resonating group method (RGM) by Kamimura \cite{Kamimura} and
within the generator coordinate method (GCM) by Uegaki
\cite{Uegaki}. Both calculations give reasonable results for the
ground state of the $^{12}$C and some excited states. However,
wave functions provided by these models are very complicated and
heavy to handle. Furthermore, although RGM ensures correct account
of nucleon exchange between different clusters, an antisymmetry
requirement on the total wave function can be violated by the
improper truncation of model space. The measure of this violation
is lacking. For example, in Ref. \cite{Kamimura} the RGM
calculation was performed with truncation to a space where the
angular momentum $l$ of $\alpha-\alpha$ relative wave function was
fixed to be equal to zero. At the same time, even the lowest
Pauli-allowed state of the $^{12}$C represents the mixture of
$l=0,\,2$ and 4. This raises the question as to whether such
truncation is consistent with the requirements of the Pauli
principle. As for the GCM, seven-dimensional numerical integrals
are involved in the calculation and the integration over generator
coordinate is replaced into summation over some mesh-points.
However, the generator coordinate is chosen to be real and its
domain is not well-defined, while only complex generator
parameters ensure the existence of inverse transition from the
generator parameter space to the coordinate space
\cite{Horiuchi_77_suppl}.

A number of macroscopic models were also applied to studying the
$^{12}$C nucleus (see, for example, Refs.
\cite{Neudatchin,Tursunov}). Within such models, clusters are
considered to be structureless particles interacting via local
potentials, which reproduce experimental $\alpha-\alpha$ phase
shifts. But the antisymmetrization prevents $\alpha-\alpha$
potential from being local. Furthermore, an accurate elimination
of the Pauli-forbidden states is needed to get any reliable
results on the structure of low-lying states of $^{12}$C nucleus.
By this reason, all macroscopic models fail to reproduce main
properties of the $3\alpha$ system.

Orthogonality condition model (OCM) has been proposed in Ref.
\cite{Saito} as an approximation of resonating group method and
applied to the investigation of the three-alpha structure  in
$^{12}$C in Refs. \cite{Horiuchi_74_51,Horiuchi_74}. In OCM the
inter-cluster wave function is required to be orthogonal to the
forbidden states and, consequently, the latter states are
completely removed from the wave function. In two-cluster systems,
OCM allows one to obtain only eigenfunctions of the
antisymmetrization operator, not eigenvalues, which enter the
Schr\"{o}dinger equation and thus affect the dynamics of cluster
system. Hence, in OCM a part of exchange effects is missing while
it could be essential, especially in three-cluster systems. This
is supported by the fact that the lack of binding energy of
$^{12}$C is observed in OCM compared to RGM calculations
\cite{alpha-nuclei}. In Ref. \cite{Horiuchi_75} Horiuchi assumed
that the $3\alpha$ redundant solutions are all reducible to
$\alpha-\alpha$ redundancy and the Pauli-allowed $3\alpha$ wave
functions can be defined from the requirement of their
orthogonality to the $\alpha-\alpha$ forbidden states. Horiuchi
also noticed that the Pauli-allowed basis functions should be
classified by the number of total oscillator quanta and indices of
the SU(3) symmetry $(\lambda,\mu)$, because the second-order
Casimir operator of the SU(3) group commutes with the operator of
permutation of the nucleon position vectors. However, most of the
three-cluster states are SU(3)-degenerate and the degree of
SU(3)-degeneration increases drastically with increasing the
number of oscillator quanta. Pauli-allowed basis functions
corresponding to the same SU(3)-symmetry indices differs in their
eigenvalues and should be labeled by additional quantum number.
But construction method of OCM does not specify the character of
the latter quantum number and, hence, in the case of
SU(3)-degeneration neither eigenvalues nor eigenfunctions of the
antisymmetrization operator can be uniquely determined.

Recently Fujiwara and co-authors proposed a new type of
three-cluster equation that employs a two-cluster RGM kernel for
the inter-cluster interaction \cite{Fujiwara_2002}. The authors
applied this equation to systems composed of three dineutrons and
three alpha-clusters. Results for binding energy for $3\alpha$
system, obtained by diagonalization of the Hamiltonian using [3]
symmetric translationary-invariant harmonic oscillator basis, have
been compared to the RGM and OCM calculations. Energy of the
$^{12}$C ground state given in Ref. \cite{Fujiwara_2002} is closer
to experimental data than the value provided by OCM, but still
underbound by 1.5 MeV compared to the $3\alpha$ RGM. The reason is
that the three-cluster equation used in Ref. \cite{Fujiwara_2002}
does not include some exchange effects contained in the genuine
three-cluster norm kernel.

The three-cluster norm kernel contains complete information about
the Pauli-allowed model space for the relative motion of clusters.
Eigenvalues and eigenfunctions of the norm kernel are solely
determined by the assumed internal wave functions of the clusters.
Hence, careful analysis of these quantities is very useful for
understanding the structure of three-cluster systems. However,
there are few papers concerning this question. Eigenvalues and
eigenfunctions of the norm kernels for $3\alpha$ and
$^{16}$O$+2\alpha$ systems have been calculated in Ref.
\cite{Kato}. They have been characterized by the SU(3)-indices and
an additional quantum number, but the meaning of the latter number
has not been established and no classification with respect to
this number has been provided. The eigenstates of the norm kernel
were expanded by the harmonic oscillator wave functions in the
rectangular coordinate representation. Only three simplest
Pauli-allowed states of the $^{24}$Mg have been given in Ref.
\cite{Kato}, while eigenfunction of the norm kernel for $^{12}$C
have not been presented. Coincidence of some eigenvalues of the
$^{24}$Mg with the eigenvalues of the two-body norm kernels of
$^{8}$Be$=\alpha+\alpha$ and $^{20}$Ne$=\alpha+^{16}$O at large
number of quanta has been noticed, but this property has not been
taken advantage of in classification of the three-cluster
Pauli-allowed states.

Summarizing, there are several problems still outstanding: construction of eigenfunctions of the
antisymmetrization operator for the three-cluster systems in simple, tractable form, complete classification of
these states (i.e., resolving SU(3)-degeneration problem) and elaboration of the truncation method of the model
space consistent with the requirements of the Pauli exclusion principle.

For the case of three-cluster systems composed of an $s$-cluster
and two neutrons we have shown that all these problems can be
resolved within the discrete representation of a complete basis of
allowed states of the multiparticle harmonic oscillator
(classified with the use of the SU(3) symmetry indices and defined
in the Fock-Bargmann space) \cite{Lashko_NPA}. Classification of
the eigenvalues of the three-cluster systems with the help of
eigenvalues of the two-body subsystem was suggested in Ref.
\cite{Lashko_NPA}. We observed that asymptotic behavior of basis
functions consistent with the requirements of the Pauli principle
gives an indication of possible decay channels of a three-cluster
nucleus and allows us to specify the most important decay
channels. Such asymptotic behavior corresponds rather to the
subsequent decay of the three-cluster system than to the so-called
"democratic decay" associated with the hyperspherical harmonics,
which are widely used for the description of three-cluster
systems.

In the present paper we extend our approach to the three-cluster
systems composed of identical $s$-clusters. In Section \ref{sec:2}
the theoretical model is briefly explained. Norm kernels of
three-fermion and three-boson systems are considered in detail in
Section \ref{sec:3} and Section \ref{sec:4}, respectively. Section
\ref{sec:5} is devoted to the analysis of the eigenvalues and
eigenfunctions of the norm kernel of the three-dineutron system.
The Pauli-allowed states of the $3\alpha$-system are constructed
and classified in Section \ref{sec:6}. The asymptotic behavior of
these functions is established  and compared to the asymptotic
form of the eigenfunctions of the $^3$H$+n+n$ norm kernel.
Concluding remarks are given in Section \ref{sec:7}. The norm
kernel of the $3\alpha$ system defined in the Fock-Bargmann space
is given in an explicit form in Appendix \ref{app:1}.

\section{Theoretical model}
\label{sec:2}

The RGM wave function of three-cluster system is represented as an
antisymmetrized product of internal cluster functions and wave
function of the relative motion of clusters. The latter is found
by solving the integro-differential equation, which is obtained
from the Schr\"{o}dinger equation upon integrating over
single-particle coordinates. The internal cluster functions are
considered to be fixed. Expanding the wave function of cluster
relative motion into the basis of the Pauli-allowed multi-particle
harmonic oscillator states $\Psi_n(\{{\bf r}\})$, one can reduce
the integro-differential equation to the set of linear algebraic
equations for the expansion coefficients. Further simplification
can be achieved by transition from the coordinate space to the
space of complex generator parameters (Fock-Bargmann space
\cite{Barg}), because the Fock-Bargmann image $\Psi_n(\{{\bf
R}\})$ of the Pauli-allowed state $\Psi_n(\{{\bf r}\})$ is much
simpler than the original. Here, $\{{\bf r}\}$ is a set of nucleon
position vectors, $\{{\bf R}\}$ is a set of generator parameters
with the help of which the dynamics of the considered degrees of
freedom will be reproduced, and $n$ is a set of quantum numbers of
basis functions.

The complete basis of the Pauli-allowed harmonic oscillator states defined both in the coordinate space and in
the Fock-Bargmann space is generated by the Slater determinant $\Phi(\{{\bf R}\},\{{\bf r}\})$ composed of the
single-particle orbitals. It is advantageous to use the modified Bloch-Brink orbitals as spatial wave functions
of the nucleons, because each orbital serves as the kernel of the transformation from the coordinate
representation to the Fock-Bargmann representation and, at the same time, it is generating function of the basis
of a harmonic oscillator. For three-cluster systems composed of three $s$-clusters containing $A_k\leq4$
nucleons, only orbitals of the following form are required:
\begin{eqnarray*}
\phi_{k}({\bf r}_i)={1\over\pi^{3/4}} \exp\left(-{1\over2}{\bf r}^2_i+\sqrt{2}({\bf R}_k\cdot{\bf r}_i)-
{1\over2}{\bf R}_k^2\right),~~i\in A_k,
\end{eqnarray*}
where complex generator parameters ${\bf R}_k$ are independent variables of the wave function $\Psi_n(\{{\bf
R}_k\})$, which describe the positions of the clusters in the Fock-Bargmann space:
\begin{eqnarray*}
{\bf R}_k={\vec{\xi}_k+i\vec{\eta}_k\over\sqrt{2}}.
\end{eqnarray*}
$\vec{\xi}_k$ and $\vec{\eta}_k$ are vectors of coordinate and momentum, respectively.

A complete basis of Pauli-allowed states in the Fock-Bargmann
space and their eigenvalues can be obtained from the overlap
integral $\bar{I}({\bf S}_k,{\bf R}_k)$ of the two Slater
determinants, which are usually defined as follows:
\begin{eqnarray*}
\bar{I}({\bf S}_k,{\bf R}_k)=\int \Phi({\bf S}_k,\{{\bf r}\})\Phi({\bf R}_k,\{{\bf r}\})d\tau.
\end{eqnarray*}
Here, integration is performed over all single-particle variables. The motion of the center of mass is eliminated
by transition from the vectors ${\bf R}_k$ to the Jacobi vectors
\begin{eqnarray*}
{\bf R}_{cm}={1\over\sqrt{A}}\left(A_1{\bf R}_1+A_2{\bf R}_2+A_3{\bf R}_3\right),
\end{eqnarray*}
\begin{eqnarray*}
{\bf a}=\sqrt{{A_1(A_2+A_3)\over A}}\left({\bf R}_1-{A_2{\bf R}_2+A_3{\bf R}_3\over {A_2+A_3}}\right), ~~{\bf
b}=\sqrt{{A_2A_3\over{A_2+A_3}}}({\bf R}_2-{\bf R}_3).
\end{eqnarray*}
Then
\begin{eqnarray*}
{\bar I}({\bf S}_k,{\bf R}_k) = \exp({\bf R}_{\rm cm} \cdot {\bf S}_{\rm cm}) \, I({\bf a},{\bf b};\tilde{\bf
a},\tilde{\bf b}),
\end{eqnarray*}
where $I({\bf a},{\bf b};\tilde{\bf a},\tilde{\bf b})$ is a translation-invariant overlap integral generally
known as the norm kernel.

The Pauli-allowed basis states $\Psi_n({\bf a},{\bf b})$ are the eigenfunctions of the norm kernel:
\begin{eqnarray*}
\Lambda_n\Psi_n({\bf a},{\bf b})=\int I({\bf a},{\bf b};\tilde{\bf a},\tilde{\bf b})\Psi_n(\tilde{\bf
a}^*,\tilde{\bf b}^*)d\mu_{\tilde{\bf a},\tilde{\bf b}},
\end{eqnarray*}
while $\Lambda_n$ are its eigenvalues. The Bargmann measure
$d\mu_{{\bf a},{\bf b}}$ is defined as follows:
\begin{eqnarray*}
d\mu_{{\bf a},{\bf b}}=\exp\{-({\bf a}{\bf a}^*)\}\,{d\vec{\xi_a}d\vec{\eta_a}\over(2\pi)^3}\,\exp\{-({\bf b}{\bf
b}^*)\}\,{d\vec{\xi_b}d\vec{\eta_b}\over(2\pi)^3}.
\end{eqnarray*}
The Hilbert-Schmidt expansion of the norm kernel
\begin{eqnarray*}
I({\bf a},{\bf b};\tilde{\bf a},\tilde{\bf b})=\sum_n\Lambda_n\Psi_n({\bf a},{\bf b})\Psi_n(\tilde{\bf
a},\tilde{\bf b})
\end{eqnarray*}
can be interpreted also as density matrix of mixed system
\cite{Landau}, with $\Lambda_n$ being the elements of the
diagonalized density matrix, which are proportional to the
realization probability of the system states defined by the
corresponding eigenfunctions.

Examine next the explicit expressions for the norm kernels of
those three-cluster systems which comprised of three identical
$s$-clusters, such as $^3n=n+n+n$, three-boson system,
$^6n=^2n+^2n+^2n$ and $^{12}$C=$\alpha+\alpha+\alpha$.

\section{Norm kernel of three-fermion system}
\label{sec:3}

In this section we consider norm kernel of three identical
fermions (for example, three nucleons with the same spin and
isospin projections), which is the main building block for the
norm kernels of three-cluster systems composed of three identical
clusters.

Instead of Jacobi vectors ${\bf a}$ and ${\bf b}$ let us introduce new complex vectors
\begin{eqnarray*}
{\bf A}={{\bf a}+i\,{\bf b}\over\sqrt{2}},\,\,\,{\bf B}={{\bf a}-i\,{\bf b}\over\sqrt{2}}.
\end{eqnarray*}
Similar transformation of creation and annihilation operator has
been shown to simplify essentially construction of a complete
bispherical harmonic oscillator basis in three-nucleon system
\cite{Moshinsky}. Further we shall demonstrate that such
transformation facilitates also construction of SU(3)-basis for
three-fermion system and, more importantly, solving eigenvalue and
eigenfunction problem for the three-cluster norm kernel.

In terms of vectors ${\bf A}$ and ${\bf B}$ the norm kernel for three identical nucleons has the following form:
\begin{eqnarray*}
I_{n+n+n}({\bf A},{\bf B};\tilde{\bf A},\tilde{\bf
B})&=&{1\over3!}\exp\left\{({\bf A}\tilde{\bf A})+({\bf
B}\tilde{\bf
B})\right\}+{1\over3!}\exp\left\{e^{i{2\pi\over3}}({\bf
A}\tilde{\bf A})+e^{-i{2\pi\over3}}({\bf B}\tilde{\bf
B})\right\}+\\
&+&{1\over3!}\exp\left\{e^{-i{2\pi\over3}}({\bf A}\tilde{\bf
A})+e^{i{2\pi\over3}}({\bf B}\tilde{\bf
B})\right\}-{1\over3!}\exp\left\{({\bf B}\tilde{\bf A})+({\bf
A}\tilde{\bf B})\right\}-\\
&-&{1\over3!}\exp\left\{e^{i{2\pi\over3}}({\bf B}\tilde{\bf
A})+e^{-i{2\pi\over3}}({\bf A}\tilde{\bf
B})\right\}-{1\over3!}\exp\left\{e^{-i{2\pi\over3}}({\bf
B}\tilde{\bf A})+e^{i{2\pi\over3}}({\bf A}\tilde{\bf B})\right\}.
\end{eqnarray*}
The latter expression can be considered as the result of action of
the antisymmetrization operator ${\hat A}$ on the first exponent,
which contains a compete basis of both the Pauli-allowed and
Pauli-forbidden harmonic oscillator states:
\begin{eqnarray*}
I_{n+n+n}({\bf A},{\bf B};\tilde{\bf A},\tilde{\bf
B})={1\over3!}\hat{A}\exp\left\{({\bf A}\tilde{\bf A})+({\bf
B}\tilde{\bf B})\right\},
\end{eqnarray*}
where $\hat{A}$ is defined in the Fock-Bargmann space as
\begin{eqnarray}
\hat{A}f({\bf A},{\bf B})&=&f({\bf A},{\bf
B})+f\left(e^{i{2\pi\over 3}}{\bf A},e^{-i{2\pi\over 3}}{\bf
B}\right)+f\left(e^{-i{2\pi\over 3}}{\bf A},e^{i{2\pi\over 3}}{\bf
B}\right)-\nonumber\\
&-&f({\bf B},{\bf A})-f\left(e^{i{2\pi\over 3}}{\bf
B},e^{-i{2\pi\over 3}}{\bf A}\right)-f\left(e^{-i{2\pi\over
3}}{\bf B},e^{i{2\pi\over 3}}{\bf A}\right). \label{antisym_3n}
\end{eqnarray}
First term corresponds to the identity permutation, second and
third terms describe two cyclic permutations, while the last three
items are associated with pair permutations and, hence, have the
minus sign. So, permutations in the system of three identical
nucleons correspond to the rotation by 120 degrees in the plane
going through all nucleons. Evidently, such rotation relates to
the transition from one Jacobi tree to another:
\begin{eqnarray*}
{\bf A}_1=e^{i{2\pi\over3}}{\bf B};~~{\bf
B}_1=e^{-i{2\pi\over3}}{\bf A}; {\bf A}_2=e^{-i{2\pi\over3}}{\bf
B};~~{\bf B}_2=e^{i{2\pi\over3}}{\bf A},
\end{eqnarray*}
where
\begin{eqnarray*}
{\bf A}_{1,2}={{\bf a}_{1,2}+i\,{\bf b}_{1,2}\over\sqrt{2}},\,\,\,{\bf B}_{1,2}={{\bf a}_{1,2}-i\,{\bf
b}_{1,2}\over\sqrt{2}}.
\end{eqnarray*}
Vectors ${\bf a}_{1,2},\,{\bf b}_{1,2}$ belong to the alternative
Jacobi trees.

Expanding the norm kernel $I_{n+n+n}\left({\bf A},{\bf B};\tilde{\bf A},\tilde{\bf B}\right)$ in powers of
vectors ${\bf A}$ and ${\bf B}$ we obtain two sums:
\begin{eqnarray*}
I_{n+n+n}\left({\bf A},{\bf B};\tilde{\bf A},\tilde{\bf B}\right)&=& \sum_{m=1}^\infty{1\over
2(m!)^2}\left\{({\bf A}\tilde{\bf A})^m ({\bf B}\tilde{\bf B})^m-({\bf B}\tilde{\bf A})^m ({\bf A}\tilde{\bf
B})^m\right\}+\\
&+&\sum_{m=0}^\infty\sum_{k=1}^\infty{1\over 2m!(m+3k)!}\left\{ ({\bf A}\tilde{\bf A})^{m+3k} ({\bf B}\tilde{\bf
B})^m+({\bf A}\tilde{\bf A})^m ({\bf B}\tilde{\bf B})^{m+3k}-\right.\\
&- &\left.({\bf B}\tilde{\bf A})^{m+3k}({\bf A}\tilde{\bf B})^m- ({\bf B}\tilde{\bf A})^m({\bf A}\tilde{\bf
B})^{m+3k}\right\}.
\end{eqnarray*}
Either of two sums generates the states with definite number of
oscillator quanta $\bar{\nu}$: the one-fold sum corresponds to
$\bar{\nu}=2m$, while the two-fold sum contains the states with
$\bar{\nu}=2m+3k.$ It should be noticed that only expressions,
which are antisymmetric with respect to nucleon permutations,
enter the above expansion of the norm kernel.

These sums each are the superpositions of partial norm kernels
$i_{(\bar{\lambda},\bar{\mu})}$ possessing definite SU(3)-symmetry
and labeled by the SU(3) indices $(\bar{\lambda},\bar{\mu})$ (see
Ref. \cite{jmp}):
\begin{eqnarray*}
{1\over 2(m!)^2}\left\{({\bf A}\tilde{\bf A})^m ({\bf B}\tilde{\bf
B})^m-({\bf B}\tilde{\bf A})^m ({\bf A}\tilde{\bf
B})^m\right\}=\sum_{\bar{\mu}=0}^m i_{(2m-2\bar{\mu},\bar{\mu})}
\left({\bf A},{\bf B};\tilde{\bf A},\tilde{\bf B}\right),
\end{eqnarray*}
\begin{eqnarray*}
{1\over 2m!(m+3k)!}\left\{({\bf A}\tilde{\bf A})^{m+3k} ({\bf
B}\tilde{\bf B})^m+({\bf A}\tilde{\bf A})^m ({\bf B}\tilde{\bf
B})^{m+3k}-({\bf B}\tilde{\bf A})^{m+3k}({\bf A}\tilde{\bf B})^m-
({\bf B}\tilde{\bf A})^m({\bf A}\tilde{\bf B})^{m+3k}\right\}=\\=
\sum_{\bar{\mu}=0}^m i_{(2m+3k-2\bar{\mu},\bar{\mu})},
\end{eqnarray*}
where
\begin{eqnarray*}
i_{(2m-2\bar{\mu},\bar{\mu})}\left({\bf A},{\bf B};\tilde{\bf
A},\tilde{\bf B}\right)&=&
{(-1)^{\bar{\mu}}(-m)_{\bar{\mu}}(-m)_{\bar{\mu}}\over2(m!)^2\bar{\mu}!(-2m+\bar{\mu}-1)_{\bar{\mu}}}
([{\bf
AB}][\tilde{\bf A}\tilde{\bf B}])^{\bar{\mu}}\times\\
& &\times\left\{({\bf A}\tilde{\bf A})^{m-\bar{\mu}}({\bf B}\tilde{\bf B})^{m-\bar{\mu}}
{_2}{F}_1\left(-m+\bar{\mu},-m+\bar{\mu};-2m+2\bar{\mu}; z\right)-\right.\\
& &-\left.(-1)^{\bar{\mu}}({\bf B}\tilde{\bf A})^{m-\bar{\mu}}({\bf A}\tilde{\bf B})^{m-\bar{\mu}}
{_2}{F}_1\left(-m+\bar{\mu},-m+\bar{\mu};-2m+2\bar{\mu}; \tilde{z}\right)\right\};
\end{eqnarray*}
\begin{eqnarray*}
i_{(2m+3k-2\bar{\mu},\bar{\mu})}\left({\bf A},{\bf B};\tilde{\bf
A},\tilde{\bf B}\right) &=&
{(-1)^{\bar{\mu}}(-m)_{\bar{\mu}}(-m-3k)_{\bar{\mu}}\over2m!(m+3k)!\bar{\mu}!(-2m-3k+\bar{\mu}-1)_{\bar{\mu}}}
([{\bf AB}][\tilde{\bf A}\tilde{\bf B}])^{\bar{\mu}}\times\\
& &\times\left\{\left\{({\bf A}\tilde{\bf A})^{m+3k-\bar{\mu}}({\bf B}\tilde{\bf B})^{m-\bar{\mu}}+ ({\bf
A}\tilde{\bf A})^{m-\bar{\mu}}({\bf B}\tilde{\bf B})^{m+3k-\bar{\mu}}\right\}\times \right.\\
& &\times{_2}{F}_1\left(-m+\bar{\mu},-m-3k+\bar{\mu};-2m-3k+2\bar{\mu}; z\right)-\\
& &-(-1)^{\bar{\mu}}\left\{({\bf B}\tilde{\bf A})^{m+3k-\bar{\mu}}({\bf A}\tilde{\bf B})^{m-\bar{\mu}}+ ({\bf
B}\tilde{\bf A})^{m-\bar{\mu}}({\bf A}\tilde{\bf B})^{m+3k-\bar{\mu}}\right\}\times\\
& &\left.\times{_2}{F}_1\left(-m+\bar{\mu},-m-3k+\bar{\mu};-2m-3k+2\bar{\mu}; \tilde{z}\right)\right\};
\end{eqnarray*}
Here $(-x)_n$ is the Pochhammer symbol, defined as follows:
\begin{eqnarray*}
(-x)_n=(-1)^n{x!\over(x-m)!};
\end{eqnarray*}
${_2}{F}_1(\alpha,\beta;\gamma;Z)$ is the hypergeometric function
with variables
\begin{eqnarray*}
z={([{\bf AB}][\tilde{\bf A}\tilde{\bf B}])\over ({\bf A}\tilde{\bf A}) ({\bf B}\tilde{\bf B})},~~\tilde{z}=
{([{\bf BA}][\tilde{\bf A}\tilde{\bf B}])\over ({\bf B}\tilde{\bf A}) ({\bf A}\tilde{\bf B})}.
\end{eqnarray*}
The number of oscillator quanta $\bar{\nu}$ relates the
SU(3)-symmetry indices $(\bar{\lambda},\bar{\mu})$ as
$\bar{\nu}=\bar{\lambda}+2\bar{\mu}.$

Partial norm kernels $i_{(\bar{\lambda},\bar{\mu})}$ are
normalized to the dimensionality of $(\bar{\lambda},\bar{\mu})$
SU(3)-representation:
\begin{eqnarray}
\int i_{(\bar{\lambda},\bar{\mu})}({\bf A},{\bf B};\tilde{\bf
A},\tilde{\bf B})d\mu_{{\bf A},{\bf
B}}={(\bar{\lambda}+1)(\bar{\mu}+1)(\bar{\lambda}+\bar{\mu}+2)\over2}.
\end{eqnarray}
It is remarkable that $i_{(2m-4\mu,2\mu)}=0.$ Hence, the latter
representation contains only the Pauli-forbidden states.
Evidently, the state corresponding to zero number of oscillator
quanta also can not be realized, i.e., $i_{(0,0)}=0.$

Basis functions characterized by zero total orbital momentum $L=0$
are generated by the $i_{(2m+3k-2\bar{\mu},\bar{\mu})}$
representations, provided that both SU(3)-indices are even:
$2m+3k=2\nu,\,\bar{\mu}=2\mu$. Explicitly, these functions are
expressible in terms of hypergeometric functions
$_2F_1(\alpha,\beta;\gamma; Z)$, with the variable
$$Z={[{\bf AB}]^2\over {\bf A}^2{\bf B}^2}.$$
Namely,
\begin{eqnarray*}
\chi^{\nu-3k-2\mu}_{(2\nu-4\mu,2\mu)}({\bf A},{\bf
B})&=&N^{\nu-3k-2\mu}_{(2\nu-4\mu,2\mu)}\left({\bf A}^{6k}-{\bf
B}^{6k}\right){\bf A}^{2n-2\mu}{\bf B}^{2n-2\mu}[{\bf A}{\bf B}]^{2\mu}\times\\
& &\times\left\{
\begin{array}{l}
{_2}F_1\left(-\nu+n+\mu,-n+\mu;-\nu+2\mu+{1\over2};Z\right),~~\nu-3k=2n;\\
({\bf
AB})\,{_2}F_1\left(-\nu+n+1+\mu,-n+\mu;-\nu+2\mu+{1\over2};Z\right),~~\nu-3k=2n+1,
\end{array}\right.
\end{eqnarray*}
\begin{eqnarray*}
N^{\nu-3k-2\mu}_{(2\nu-4\mu,2\mu)}&=&{1\over\sqrt{2(\nu+3k-2\mu)!(\nu-3k-2\mu)!}}\times\\
& &\times \sqrt{(2\nu+1-4\mu)!\over(2\mu)!(2\nu+1-2\mu)!}
\sqrt{{(2\nu-2\mu)!!(2\nu-4\mu-1)!!(2\mu-1)!!\over(2\nu-2\mu+1)!!(2\nu-4\mu)!!(2\mu)!!}}.
\end{eqnarray*}
A set of quantum numbers of functions
$\chi^{\nu-3k-2\mu}_{(2\nu-4\mu,2\mu)}({\bf A},{\bf B})$ includes
the total number of oscillator quanta $2\nu,$ the SU(3) symmetry
indices $(\lambda,\mu)$, and additional quantum number $k$ for the
SU(3)-degenerated states\footnote{An upper index of basis
functions corresponds to the minimal of the number of quanta along
${\bf A}$ and ${\bf B}$}. Total orbital momentum $L$ and its
projection $M$ are equal to zero and, hence, will be omitted in
what follows.

Functions $\chi^{\nu-3k-2\mu}_{(2\nu-4\mu,2\mu)}({\bf A},{\bf B})$
comply with requirements of the Pauli exclusion principle. It is
easy to verify this acting on these functions by
antisymmetrization operator $\hat{A}$ defined in
(\ref{antisym_3n}).
\begin{eqnarray*}
\hat{A}\chi^{\nu-3k-2\mu}_{(2\nu-4\mu,2\mu)}({\bf A},{\bf
B})=\Lambda_{(2\nu-4\mu,2\mu)_{k}}\chi^{\nu-3k-2\mu}_{(2\nu-4\mu,2\mu)},~~
\Lambda_{(2\nu-4\mu,2\mu)_{k}}=
\left\{%
\begin{array}{ll}
    1, & \hbox{$1\leq k\leq\left[{\nu-2\mu\over3}\right]$;} \\
    0, & \hbox{$k=0$.} \\
\end{array}%
\right.
\end{eqnarray*}
Here $[x]$ denotes an integer part of the number $x$.

Hence, at each value of the number of oscillator quanta $\nu$
there are $\left[{\nu-2\mu\over3}\right]$ Pauli-allowed basis
functions and one Pauli-forbidden state belonging to the same
SU(3) representation $(2\nu-4\mu,2\mu).$ All the Pauli-allowed
states have unit eigenvalues, while the Pauli-forbidden states
correspond to zero eigenvalues. This implies that a norm kernel of
three-nucleon system can be treated as the density matrix of the
pure system. Indeed, in this case dynamics of all degrees of
freedom is considered and no averaging is performed.

In representation of ${\bf A}$ and ${\bf B}$ vectors an
SU(3)-projected norm kernel $i_{(2\nu-4\mu,2\mu)}({\bf A},{\bf
B};\tilde{\bf A},\tilde{\bf B})$ has a diagonal form:
\begin{eqnarray*}
i_{(2\nu-4\mu,2\mu)}({\bf A},{\bf B};\tilde{\bf A},\tilde{\bf
B})=\sum_{k=1}^{\left[{\nu-2\mu\over3}\right]}\chi^{\nu-3k-2\mu}_{(2\nu-4\mu,2\mu)}({\bf
A},{\bf B})\chi^{\nu-3k-2\mu}_{(2\nu-4\mu,2\mu)}(\tilde{\bf
A},\tilde{\bf B}).
\end{eqnarray*}
Hence, solving the eigenvalue and eigenvectors problem for
three-nucleon system becomes a trivial procedure. It is also easy
to distinguish basis states belonging to the SU(3)-degenerated
representation by the additional quantum number $k,$ which
explicitly enters the expression for the basis functions. However,
due to the fact that all eigenvalues of the Pauli-allowed states
equal to unity, eigenfunctions of the norm kernel are not uniquely
determined. Any unitary transformation applied to the functions
$\chi^{\nu-3k-2\mu}_{(2\nu-4\mu,2\mu)}({\bf A},{\bf B})$ gives a
new set of basis functions, which are also antisymmetric with
respect to the nucleon permutation. For example, hyperspherical
harmonics can be successfully applied to studying three-nucleon
systems.

\section{Norm kernel of three-boson system}
\label{sec:4}

In this section we consider norm kernel of three identical bosons
$I_{3b}({\bf A},{\bf B};\tilde{\bf A},\tilde{\bf B})$, which
generates basis functions needed for the description of
three-cluster systems composed of three identical boson clusters,
such as $^6n=^2n+^2n+^2n$ and $^{12}$C=$\alpha+\alpha+\alpha$.

In terms of vectors ${\bf A}$ and ${\bf B}$ the norm kernel for
three identical bosons is very similar to that of three fermions:
\begin{eqnarray*}
I_{3b}({\bf A},{\bf B};\tilde{\bf A},\tilde{\bf
B})&=&{1\over3!}\exp\left\{({\bf A}\tilde{\bf A})+({\bf
B}\tilde{\bf
B})\right\}+{1\over3!}\exp\left\{e^{i{2\pi\over3}}({\bf
A}\tilde{\bf A})+e^{-i{2\pi\over3}}({\bf B}\tilde{\bf
B})\right\}+\\
&+&{1\over3!}\exp\left\{e^{-i{2\pi\over3}}({\bf A}\tilde{\bf
A})+e^{i{2\pi\over3}}({\bf B}\tilde{\bf
B})\right\}+{1\over3!}\exp\left\{({\bf B}\tilde{\bf A})+({\bf
A}\tilde{\bf B})\right\}+\\
&+&{1\over3!}\exp\left\{e^{i{2\pi\over3}}({\bf B}\tilde{\bf
A})+e^{-i{2\pi\over3}}({\bf A}\tilde{\bf
B})\right\}+{1\over3!}\exp\left\{e^{-i{2\pi\over3}}({\bf
B}\tilde{\bf A})+e^{i{2\pi\over3}}({\bf A}\tilde{\bf B})\right\}.
\end{eqnarray*}
The only difference between the norm kernels of three fermions and
three bosons is that the latter is symmetric with respect to
nucleon permutation, while the former is antisymmetric. Naturally,
$I_{3b}({\bf A},{\bf B};\tilde{\bf A},\tilde{\bf B})$ can be
considered as the result of action of the symmetrization operator
$\hat{P}$ on the first exponent:
\begin{eqnarray*}
I_{3b}({\bf A},{\bf B};\tilde{\bf A},\tilde{\bf
B})={1\over3!}\hat{P}\exp\left\{({\bf A}\tilde{\bf A})+({\bf
B}\tilde{\bf B})\right\},
\end{eqnarray*}
where $\hat{P}$ is defined in the Fock-Bargmann space as
\begin{eqnarray}
\hat{P}f({\bf A},{\bf B})&=&f({\bf A},{\bf
B})+f\left(e^{i{2\pi\over 3}}{\bf A},e^{-i{2\pi\over 3}}{\bf
B}\right)+f\left(e^{-i{2\pi\over 3}}{\bf A},e^{i{2\pi\over 3}}{\bf
B}\right)+\nonumber\\
&+&f({\bf B},{\bf A})+f\left(e^{i{2\pi\over 3}}{\bf
B},e^{-i{2\pi\over 3}}{\bf A}\right)+f\left(e^{-i{2\pi\over
3}}{\bf B},e^{i{2\pi\over 3}}{\bf A}\right). \label{sym_3b}
\end{eqnarray}
Meaning of each term is the same as in Eq. (\ref{antisym_3n}).

Expansion of the norm kernel $I_{3b}\left({\bf A},{\bf
B};\tilde{\bf A},\tilde{\bf B}\right)$ in powers of vectors ${\bf
A}$ and ${\bf B}$ gives the result:
\begin{eqnarray*}
I_{3b}\left({\bf A},{\bf B};\tilde{\bf A},\tilde{\bf B}\right)&=&
\sum_{m=1}^\infty{1\over 2(m!)^2}\left\{({\bf A}\tilde{\bf A})^m
({\bf B}\tilde{\bf B})^m+({\bf B}\tilde{\bf A})^m ({\bf
A}\tilde{\bf
B})^m\right\}+\\
&+&\sum_{m=0}^\infty\sum_{k=1}^\infty{1\over 2m!(m+3k)!}\left\{
({\bf A}\tilde{\bf A})^{m+3k} ({\bf B}\tilde{\bf
B})^m+({\bf A}\tilde{\bf A})^m ({\bf B}\tilde{\bf B})^{m+3k}+\right.\\
&+ &\left.({\bf B}\tilde{\bf A})^{m+3k}({\bf A}\tilde{\bf B})^m+
({\bf B}\tilde{\bf A})^m({\bf A}\tilde{\bf B})^{m+3k}\right\}.
\end{eqnarray*}
Of course, only expressions, which are symmetric with respect to
boson permutation, enter the above expansion of the norm kernel.
Notice, that the above-mentioned symmetry results in essential
reduction of the basis states, which can be realized in
three-boson norm kernel. Indeed, the first exponent
$\exp\left\{({\bf A}\tilde{\bf A})+({\bf B}\tilde{\bf
B})\right\}$, which generates both the Pauli-allowed and the
Pauli-forbidden basis functions, contains six times more states
than $I_{3b}\left({\bf A},{\bf B};\tilde{\bf A},\tilde{\bf
B}\right)$ norm kernel.

By analogy with three-fermion system, each sum corresponds to the
fixed number of quanta $\bar{\nu}$ and can be presented as a sum
of partial norm kernels $\bar{i}_{(\bar{\lambda},\bar{\mu})}$
belonging to the SU(3)-representations
$(\bar{\lambda},\bar{\mu}):$
\begin{eqnarray*}
{1\over 2(m!)^2}\left\{({\bf A}\tilde{\bf A})^m ({\bf B}\tilde{\bf
B})^m+({\bf B}\tilde{\bf A})^m ({\bf A}\tilde{\bf
B})^m\right\}=\sum_{\bar{\mu}=0}^m
\bar{i}_{(2m-2\bar{\mu},\bar{\mu})} \left({\bf A},{\bf
B};\tilde{\bf A},\tilde{\bf B}\right),
\end{eqnarray*}
\begin{eqnarray*}
{1\over 2m!(m+3k)!}\left\{({\bf A}\tilde{\bf A})^{m+3k} ({\bf
B}\tilde{\bf B})^m+({\bf A}\tilde{\bf A})^m ({\bf B}\tilde{\bf
B})^{m+3k}+({\bf B}\tilde{\bf A})^{m+3k}({\bf A}\tilde{\bf B})^m+
({\bf B}\tilde{\bf A})^m({\bf A}\tilde{\bf B})^{m+3k}\right\}=\\=
\sum_{\bar{\mu}=0}^m \bar{i}_{(2m+3k-2\bar{\mu},\bar{\mu})},
\end{eqnarray*}
where
\begin{eqnarray*}
\bar{i}_{(2m-2\bar{\mu},\bar{\mu})}\left({\bf A},{\bf
B};\tilde{\bf A},\tilde{\bf B}\right)&=&
{(-1)^{\bar{\mu}}(-m)_{\bar{\mu}}(-m)_{\bar{\mu}}\over2(m!)^2\bar{\mu}!(-2m+\bar{\mu}-1)_{\bar{\mu}}}
([{\bf
AB}][\tilde{\bf A}\tilde{\bf B}])^{\bar{\mu}}\times\\
& &\times\left\{({\bf A}\tilde{\bf A})^{m-\bar{\mu}}({\bf
B}\tilde{\bf B})^{m-\bar{\mu}}
{_2}{F}_1\left(-m+\bar{\mu},-m+\bar{\mu};-2m+2\bar{\mu}; z\right)+\right.\\
& &+\left.(-1)^{\bar{\mu}}({\bf B}\tilde{\bf
A})^{m-\bar{\mu}}({\bf A}\tilde{\bf B})^{m-\bar{\mu}}
{_2}{F}_1\left(-m+\bar{\mu},-m+\bar{\mu};-2m+2\bar{\mu};
\tilde{z}\right)\right\};
\end{eqnarray*}
\begin{eqnarray*}
\bar{i}_{(2m+3k-2\bar{\mu},\bar{\mu})}\left({\bf A},{\bf
B};\tilde{\bf A},\tilde{\bf B}\right) &=&
{(-1)^{\bar{\mu}}(-m)_{\bar{\mu}}(-m-3k)_{\bar{\mu}}\over2m!(m+3k)!\bar{\mu}!(-2m-3k+\bar{\mu}-1)_{\bar{\mu}}}
([{\bf AB}][\tilde{\bf A}\tilde{\bf B}])^{\bar{\mu}}\times\\
& &\times\left\{\left\{({\bf A}\tilde{\bf
A})^{m+3k-\bar{\mu}}({\bf B}\tilde{\bf B})^{m-\bar{\mu}}+ ({\bf
A}\tilde{\bf A})^{m-\bar{\mu}}({\bf B}\tilde{\bf B})^{m+3k-\bar{\mu}}\right\}\times \right.\\
& &\times{_2}{F}_1\left(-m+\bar{\mu},-m-3k+\bar{\mu};-2m-3k+2\bar{\mu}; z\right)+\\
& &+(-1)^{\bar{\mu}}\left\{({\bf B}\tilde{\bf
A})^{m+3k-\bar{\mu}}({\bf A}\tilde{\bf B})^{m-\bar{\mu}}+ ({\bf
B}\tilde{\bf A})^{m-\bar{\mu}}({\bf A}\tilde{\bf B})^{m+3k-\bar{\mu}}\right\}\times\\
&
&\left.\times{_2}{F}_1\left(-m+\bar{\mu},-m-3k+\bar{\mu};-2m-3k+2\bar{\mu};
\tilde{z}\right)\right\}.
\end{eqnarray*}
Arguments $z$ and $\tilde{z}$ of the hypergeometric functions are
the same as those in the previous section.

In contrast to three-fermion system,
$\bar{i}_{(2m-4\mu,2\mu)}\neq0,$ while
$\bar{i}_{(2m-4\mu-2,2\mu+1)}=0.$ Therefore, for three-boson
system all the states belonging to $(2\nu-4\mu,2\mu)$
representation are allowed by the Pauli principle. Basis functions
characterized by zero total orbital momentum $L=0$ are of the
following form:
\begin{eqnarray}
\psi^{\nu-3k-2\mu}_{(2\nu-4\mu,2\mu)}({\bf A},{\bf
B})&=&\bar{N}^{\nu-3k-2\mu}_{(2\nu-4\mu,2\mu)}\left({\bf
A}^{6k}+{\bf
B}^{6k}\right){\bf A}^{2n-2\mu}{\bf B}^{2n-2\mu}[{\bf A}{\bf B}]^{2\mu}\times \nonumber\\
& &\times\left\{
\begin{array}{l}
{_2}F_1\left(-\nu+n+\mu,-n+\mu;-\nu+2\mu+{1\over2};Z\right),~~\nu-3k=2n;\\
({\bf
AB})\,{_2}F_1\left(-\nu+n+1+\mu,-n+\mu;-\nu+2\mu+{1\over2};Z\right),~~\nu-3k=2n+1,
\end{array}\right.
\label{boson_func}
\end{eqnarray}
\begin{eqnarray*}
\bar{N}^{\nu-3k-2\mu}_{(2\nu-4\mu,2\mu)}&=&{1\over\sqrt{2(1+\delta_{k,0})(\nu+3k-2\mu)!(\nu-3k-2\mu)!}}\times\\
& &\times \sqrt{(2\nu+1-4\mu)!\over(2\mu)!(2\nu+1-2\mu)!}
\sqrt{{(2\nu-2\mu)!!(2\nu-4\mu-1)!!(2\mu-1)!!\over(2\nu-2\mu+1)!!(2\nu-4\mu)!!(2\mu)!!}}.
\end{eqnarray*}
Functions $\psi^{\nu-3k-2\mu}_{(2\nu-4\mu,2\mu)}({\bf A},{\bf B})$
are the eigenfunctions of the symmetrization operator ${\hat P}$
defined in (\ref{sym_3b}):
\begin{eqnarray*}
\hat{P}\psi^{\nu-3k-2\mu}_{(2\nu-4\mu,2\mu)}({\bf A},{\bf
B})=\psi^{\nu-3k-2\mu}_{(2\nu-4\mu,2\mu)},~~ 0\leq
k\leq\left[{\nu-2\mu\over3}\right].
\end{eqnarray*}
Finally, at each value of the number of oscillator quanta
 $\left[{\nu-2\mu\over3}\right]+1$ basis
functions, belonging to  SU(3)-representation $(2\nu-4\mu,2\mu)$
and having zero orbital momentum, can be constructed, and all
these states are consistent with the requirements of the Pauli
principle.

In the next section we will demonstrate that the Pauli-allowed
states of three-cluster systems composed of identical boson
clusters can be represented as a superposition of eigenfunctions
$\psi^{\nu-3k-2\mu}_{(2\nu-4\mu,2\mu)}({\bf A},{\bf B})$ of
three-boson norm kernel. Indeed, the structure of the latter
functions reflects the symmetry with respect to the permutation of
clusters as a whole and, in addition, these functions are the
eigenfunctions of the second-order Casimir operator. Further it
will be shown that using
$\psi^{\nu-3k-2\mu}_{(2\nu-4\mu,2\mu)}({\bf A},{\bf B})$ functions
we essentially simplify the eigenvalue and eigenfunction problem
for three-cluster systems.

\section{Norm kernel of three dineutrons}
\label{sec:5}

Norm kernel of three-dineutron system $I_{3d'}$ can be obtained by
squaring of the three-neutron norm kernel $I_{3n}$:
\begin{eqnarray*}
I_{3d'}\left({\bf A},{\bf B};\tilde{\bf A},\tilde{\bf
B}\right)=3!\left[I_{3n}\left({{\bf A}\over\sqrt{2}}\,,{{\bf
B}\over\sqrt{2}}\,;{\tilde{\bf A}\over\sqrt{2}}\,,{\tilde{\bf
B}\over\sqrt{2}}\right)\right]^2.
\end{eqnarray*}
Replacement
\begin{eqnarray*}
{\bf A}\rightarrow{1\over\sqrt{2}}\,{\bf A},~~{\bf
B}\rightarrow{1\over\sqrt{2}}\,{\bf B},~~\tilde{\bf
A}\rightarrow{1\over\sqrt{2}}\,\tilde{\bf A},~~\tilde{\bf
B}\rightarrow{1\over\sqrt{2}}\,\tilde{\bf B}
\end{eqnarray*}
is required to ensure correct elimination of the center-of-mass
motion in $3d'$ system. Such scaling transformation of the Jacobi
vectors accounts for the fact that each cluster in the $3d'$
system consists of two neutrons.

The norm kernel $I_{3d'}$ can be presented as a three-termed
antisymmetrization operator acting on the exponent, which contains
a complete basis of both the Pauli-allowed and the Pauli-forbidden
states:
\begin{eqnarray*}
I_{3d'}({\bf A},{\bf B};\tilde{\bf A},\tilde{\bf
B})={1\over3!}(\hat{A}_1+\hat{A}_2+\hat{A}_3)\exp\left\{({\bf
A}\tilde{\bf A})+({\bf B}\tilde{\bf B})\right\}.
\end{eqnarray*}
Each of three above-mentioned operators combines certain group of
permutations.

The first operator $\hat{A}_1$ coincides with the symmetrization
operator $\hat{P}$ of three-boson system defined in
(\ref{sym_3b}). Hence, $\hat{A}_1=\hat{P}$ corresponds to the
permutations of dineutron clusters as a whole.

The second part of antisymmetrization operator is defined as
\begin{eqnarray}
\hat{A}_2f({\bf A},{\bf B}) &=& 2\left\{f\left(-{{\bf
A}\over2},-{{\bf B}\over2}\right)+f\left(-e^{i{2\pi\over 3}}{{\bf
A}\over2},-e^{-i{2\pi\over 3}}{{\bf B}\over2}\right)+
f\left(-e^{-i{2\pi\over 3}}{{\bf
A}\over2},-e^{i{2\pi\over 3}}{{\bf B}\over2}\right)\right.+\nonumber\\
 &+&\left.f\left(-{{\bf
B}\over2},-{{\bf A}\over2}\right)+f\left(-e^{i{2\pi\over 3}}{{\bf
B}\over2},-e^{-i{2\pi\over 3}}{{\bf
A}\over2}\right)+f\left(-e^{-i{2\pi\over 3}}{{\bf
B}\over2},-e^{i{2\pi\over 3}}{{\bf A}\over2}\right)\right\}.
\label{antisym2_3d}
\end{eqnarray}
It is associated with permutations of nucleons between all three
dineutron clusters and includes simultaneous pair or cyclic
permutations of spin-up and spin-down neutrons. Notice that only
the first term of operator $\hat{A}_2$ results in a rearrangement
of clusters due to the permutation of identical neutrons.
Remaining five terms can be obtained by permutation of clusters as
a whole following the rearrangement associated with the first
term.

Finally, the last operator $\hat{A}_3$ is a little bit more
complicated:
\begin{eqnarray*}
\hat{A}_3=\hat{A}_3^0+\hat{A}_3^++\hat{A}_3^-,
\end{eqnarray*}
where

\begin{eqnarray}
\hat{A}_3^0f({\bf A},{\bf B}) &=& -2\left\{f\left({{\bf A}+{\bf
B}\over2},{{\bf A}+{\bf B}\over2}\right)+f\left(e^{i{2\pi\over
3}}{{\bf A}+{\bf B}\over2},e^{-i{2\pi\over 3}}{{\bf A}+{\bf
B}\over2}\right)+\right.\nonumber\\
& +&\left. f\left(e^{-i{2\pi\over 3}}{{\bf A}+{\bf
B}\over2},e^{i{2\pi\over 3}}{{\bf A}+{\bf
B}\over2}\right)\right\}; \nonumber\\
\hat{A}_3^+f({\bf A},{\bf B})
&=&-2\left\{f\left({1\over2}\left[e^{i{2\pi\over 3}}{\bf
A}+e^{-i{2\pi\over 3}}{\bf B}\right],{1\over2}\left[e^{i{2\pi\over
3}}{\bf
A}+e^{-i{2\pi\over 3}}{\bf B}\right]\right) \right.\nonumber\\
&+&\left.
f\left({e^{i{2\pi\over 3}}\over2}\left[e^{i{2\pi\over 3}}{\bf
A}+e^{-i{2\pi\over 3}}{\bf B}\right],{e^{-i{2\pi\over
3}}\over2}\left[e^{i{2\pi\over 3}}{\bf A}+e^{-i{2\pi\over 3}}{\bf
B}\right]\right)+\right.\nonumber\\
 &+&\left.
f\left({e^{-i{2\pi\over 3}}\over2}\left[e^{i{2\pi\over 3}}{\bf
A}+e^{-i{2\pi\over 3}}{\bf B}\right],{e^{i{2\pi\over
3}}\over2}\left[e^{i{2\pi\over 3}}{\bf A}+e^{-i{2\pi\over 3}}{\bf
B}\right]\right)\right\}; \nonumber\\
\hat{A}_3^-f({\bf A},{\bf B})&=&-2\left\{
f\left({1\over2}\left[e^{-i{2\pi\over 3}}{\bf A}+e^{i{2\pi\over
3}}{\bf B}\right],{1\over2}\left[e^{-i{2\pi\over 3}}{\bf
A}+e^{i{2\pi\over 3}}{\bf
B}\right]\right)+\right.\nonumber\\
 &+&\left.
f\left({e^{i{2\pi\over 3}}\over2}\left[e^{-i{2\pi\over 3}}{\bf
A}+e^{i{2\pi\over 3}}{\bf B}\right],{e^{-i{2\pi\over
3}}\over2}\left[e^{-i{2\pi\over 3}}{\bf A}+e^{i{2\pi\over 3}}{\bf
B}\right]\right)+\right.\nonumber\\
 &+&\left.
f\left({e^{-i{2\pi\over 3}}\over2}\left[e^{-i{2\pi\over 3}}{\bf
A}+e^{i{2\pi\over 3}}{\bf B}\right],{e^{i{2\pi\over
3}}\over2}\left[e^{-i{2\pi\over 3}}{\bf A}+e^{i{2\pi\over 3}}{\bf
B}\right]\right)
 \right\}.
\label{antisym3_3d}
\end{eqnarray}
Operator $\hat{A}_3$ combines permutations of nucleons between two
clusters, with the third cluster being a spectator. In this case
the spin-up neutrons undergo cyclic permutation, while the
spin-down neutrons experience pair permutation and vice versa.
Depending on which cluster is a spectator, we obtain operators
$\hat{A}_3^0,\hat{A}_3^{+}$ or $\hat{A}_3^{-}.$ The three-termed
structure of each operator corresponds to three cyclic
permutations of clusters as a whole.

It is natural to seek the eigenfunctions of the $3d'$ norm kernel
in the form of superposition of the eigenfunctions of the
three-boson norm kernel, because the dineutron clusters should
obey boson statistics. Hence, the Pauli-allowed basis functions of
the three-dineutron system $\Psi_{(2\nu-4\mu,2\mu)_{k}}({\bf
A},{\bf B})$ characterized by positive parity, zero values of
total orbital momentum and spin $L=S=0$ and belonging to the
$(2\nu-4\mu,2\mu)$ representation of the SU(3)-group can be
written as:
\begin{eqnarray}
\label{wf_expand} \Psi_{(2\nu-4\mu,2\mu)_{k}}({\bf A},{\bf
B})=\sum_{q=0}^{[{\nu-2\mu\over3}]}D^{\nu-3q-2\mu}_{(2\nu-4\mu,2\mu)_{k}}\psi_{(2\nu-4\mu,2\mu)}^{\nu-3q-2\mu}({\bf
A},{\bf B}).
\end{eqnarray}
Expansion coefficients $D^{\nu-3q-2\mu}_{(2\nu-4\mu,2\mu)_{k}}$,
along with eigenvalues $\Lambda_{(2\nu-4\mu,2\mu)_{k}}$
corresponding to the Pauli-allowed states
$\Psi_{(2\nu-4\mu,2\mu)_{k}}({\bf A},{\bf B})$, can be found by
diagonalization of the three-dineutron system norm kernel
$i^{3d'}_{(2\nu-4\mu,2\mu)}$ projected on the states with the
fixed number of quanta and definite SU(3)-symmetry:
\begin{eqnarray}
i^{3d'}_{(2\nu-4\mu,2\mu)}&=&\sum_{q=0}^{[{\nu-2\mu\over3}]}\sum_{\tilde{q}=0}^{[{\nu-2\mu\over3}]-q}
\langle2\nu,\nu-3q-2\mu|2\nu,\nu-3q-3\tilde{q}-2\mu\rangle_{3d'}\times\nonumber\\
& &\times
\left(\psi_{(2\nu-4\mu,2\mu)}^{\nu-3q-2\mu}\tilde{\psi}_{(2\nu-4\mu,2\mu)}^{\nu-3q-3\tilde{q}-2\mu}+
\psi_{(2\nu-4\mu,2\mu)}^{\nu-3q-3\tilde{q}-2\mu}\tilde{\psi}_{(2\nu-4\mu,2\mu)}^{\nu-3q-2\mu}\right).
\label{part_norm_6n}
\end{eqnarray}
In order to construct characteristic equation, let us consider the
action of the antisymmetrization operator
$\hat{A}=\hat{A}_1+\hat{A}_2+\hat{A}_3$ on the functions
$\psi_{(2\nu-4\mu,2\mu)}^{\nu-3q-2\mu}.$

It is easy to verify that
\begin{eqnarray*}
{1\over3!}\hat{A}_1\,\psi_{(2\nu-4\mu,2\mu)}^{\nu-3q-2\mu}({\bf
A}, {\bf B})=\psi_{(2\nu-4\mu,2\mu)}^{\nu-3q-2\mu}({\bf A}, {\bf
B}),~~{1\over3!}\hat{A}_2\,\psi_{(2\nu-4\mu,2\mu)}^{\nu-3q-2\mu}({\bf
A}, {\bf
B})={1\over2^{2\nu-1}}\,\psi_{(2\nu-4\mu,2\mu)}^{\nu-3q-2\mu}({\bf
A}, {\bf B}).
\end{eqnarray*}
Notice that all terms entering the expressions specifying
operators $\hat{A}_1$ or $\hat{A}_2$ give the same result acting
on basis functions $\psi_{(2\nu-4\mu,2\mu)}^{\nu-3q-2\mu}$.

The last part of antisymmetrization operator provides nonzero
result only for $(2\nu,0)$ representation:
\begin{eqnarray*}
{1\over3!}\hat{A}_3\,\psi_{(2\nu-4\mu,2\mu)}^{\nu-3q-2\mu}({\bf
A}, {\bf B})&=&-
\delta_{\mu,0}{6\over4^\nu}\sum_{\tilde{q}=0}^{[\nu/3]-q}\psi_{(2\nu-4\mu,2\mu)}^{\nu-3q-3\tilde{q}-2\mu}({\bf
A}, {\bf B})\,{1\over\sqrt{(1+\delta_{q,0})(1+\delta_{\tilde{q},-q})}}\times\\
&\times &
{(2\nu)!\over\sqrt{(\nu+3q+3\tilde{q})!(\nu-3q-3\tilde{q})!(\nu+3q)!(\nu-3q)!}}.
\end{eqnarray*}

Consequently, the matrix elements of the secular equation look
like
\begin{eqnarray*}
\langle2\nu,\nu-3q-2\mu|2\nu,\nu-3q-3\tilde{q}-2\mu\rangle_{3d'}&=&\left(1+{1\over2^{2\nu-1}}\right)\delta_{\tilde{q},0}-
\delta_{\mu,0}{6\over4^\nu}{1\over\sqrt{(1+\delta_{q,0})(1+\delta_{\tilde{q},-q})}}\times\\
&\times &
{(2\nu)!\over\sqrt{(\nu+3q+3\tilde{q})!(\nu-3q-3\tilde{q})!(\nu+3q)!(\nu-3q)!}}.
\end{eqnarray*}
The action of the antisymmetrization operator ${\hat A}$ on the
basis functions $\psi_{(2\nu-4\mu,2\mu)}^{\nu-3q-2\mu}$ results
only in diagonal matrix elements of the characteristic equation
for $\mu\neq0$:
\begin{eqnarray*}
\hat{A}\psi^{\nu-3k-2\mu}_{(2\nu-4\mu,2\mu)}({\bf A},{\bf
B})=\left(1+{1\over2^{2\nu-1}}\right)\psi^{\nu-3k-2\mu}_{(2\nu-4\mu,2\mu)}({\bf
A},{\bf B}),~~\mu\neq0.
\end{eqnarray*}
By this reason, the Pauli-allowed basis functions of the $3d'$
system belonging to the SU(3) representation $(2\nu-4\mu,2\mu)$
with $\mu\neq0$ coincide with the eigenfunctions of the
three-boson norm kernel and have identical, although not equal to
unity, eigenvalues $\Lambda_{(2\nu-4\mu,2\mu)_k}$:
\begin{eqnarray*}
\Psi_{(2\nu-4\mu,2\mu)_{k}}({\bf A},{\bf
B})=\psi_{(2\nu-4\mu,2\mu)}^{\nu-3k-2\mu}({\bf A},{\bf
B}),~~\Lambda_{(2\nu-4\mu,2\mu)_{k}}=1+{1\over2^{2\nu-1}},~~ 0\leq
k\leq\left[{\nu-2\mu\over3}\right],~~\mu\neq0.
\end{eqnarray*}
The Pauli-allowed states of $(2\nu,0)$ representations are
characterized by the same eigenvalues at a given number of
oscillator quanta $\nu$. However, there is also one
Pauli-forbidden state, corresponding to zero eigenvalue:
\begin{eqnarray*}
\Lambda_{(2\nu,0)}=(1-\delta_{k,0})\left(1+{1\over2^{2\nu-1}}\right),~~
0\leq k\leq\left[{\nu\over3}\right].
\end{eqnarray*}
Eigenfunctions $\Psi_{(2\nu,0)_{k}}({\bf A},{\bf B})$ of the
antisymmetrization operator have more complicated form, being a
superposition of functions $\psi^{\nu-3q}_{(2\nu,0)}$ (see Eq.
(\ref{wf_expand})). However, with increasing the number of
oscillator quanta the Pauli-forbidden state
$\Psi_{(2\nu,0)_{k=0}}({\bf A},{\bf B})$ takes simple analytical
form:
\begin{eqnarray*}
\Psi_{(2\nu,0)_{k=0}}&\rightarrow&{1\over\sqrt{3}}{1\over\sqrt{(2\nu+1)!}}\left({\bf
a}^{2\nu}+ {\bf a}_1^{2\nu}+{\bf a}_2^{2\nu}\right)=\\
&=&\sum_{q=0}^{[\nu/3]}\sqrt{6\over1+\delta_{q,0}}\,\left({1\over2}\right)^{\nu}\sqrt{{(2\nu)!\over(\nu-3q)!(\nu+3q)!}}\psi^{\nu-3q}_{(2\nu,0)}({\bf
A},{\bf B}).
\end{eqnarray*}
It is quite obvious that such asymptotic behavior corresponds to
the case when a dineutron moves away from the $^4n=^2n+^2n$
subsystem (tetraneutron), with the latter being in the
Pauli-forbidden state.

Similar behavior is peculiar also for the Pauli-allowed states. By
means of the unitary transformation the eigenfunctions of the
$3d'$ norm kernel corresponding to nonzero eigenvalues can be
rearranged so that
\begin{eqnarray}
\label{func_as}
\Psi_{(2\nu-4\mu,2\mu)_{k}}\rightarrow\Psi_{(2\nu-4\mu,2\mu)_{k}}^{\rm
as}&=&{1\over\sqrt{3}}\left(\phi_{(2\nu-4\mu,2\mu)}^{2k-2\mu}({\bf
a},{\bf b})+\phi_{(2\nu-4\mu,2\mu)}^{2k-2\mu}({\bf a}_1,{\bf
b}_1)+\phi_{(2\nu-4\mu,2\mu)}^{2k-2\mu}({\bf a}_2,{\bf
b}_2)\right)=
\nonumber\\
&=&\sum_{q=0}^{[{\nu-2\mu\over3}]}
\sqrt{6\over1+\delta_{q,0}}\,\bar{{\cal
K}}^{1/2}_{2k}(\nu-3q-2\mu)\psi^{\nu-3q-2\mu}_{(2\nu-4\mu,2\mu)}({\bf
A},{\bf B}).
\end{eqnarray}
Here functions $\phi_{(2\nu-4\mu,2\mu)}^{k-2\mu}({\bf a},{\bf b})$
are defined as
\begin{eqnarray*}
\phi^{2m-2\mu}_{(2\nu-4\mu,2\mu)}({\bf a},{\bf
b})&=&N^{2m-2\mu}_{(2\nu-4\mu,2\mu)} [{\bf ab}]^{2\mu}{\bf
a}^{2\nu-2m-2\mu}{\bf
b}^{2m-2\mu}\times\\
&
&\times{_2}F_1\left(-\nu+m+\mu,-m+\mu;-\nu+2\mu+{1\over2};Z\right),~~Z={[{\bf
ab}]^2\over {\bf a}^2{\bf b}^2}
\end{eqnarray*}
with $N^{2m-2\mu}_{(2\nu-4\mu,2\mu)}$ being a normalization
coefficient:
\begin{eqnarray*}
N^{2m-2\mu}_{(2\nu-4\mu,2\mu)}=
\sqrt{(2\nu+1-4\mu)!(2\nu-2\mu)!!(2\nu-4\mu-1)!!(2\mu-1)!!\over
(2\mu)!(2\nu+1-2\mu)!(2\nu-2m-2\mu)!(2m-2\mu)!(2\nu-2\mu+1)!!(2\nu-4\mu)!!(2\mu)!!}.
\end{eqnarray*}
$\bar{{\cal K}}^{1/2}_{2k}(\nu-3q-2\mu)$ are the so-called
Kravchuk polynomials of a discrete variable \cite{Kravchuk}, which
are specified and orthogonal on the interval $2\mu\leq
m\leq2\nu-2\mu$:
\begin{eqnarray}
\label{Kravchuk} \bar{{\cal
K}}^{1/2}_{2k}(\nu-3q-2\mu)=\left({1\over2}\right)^{\nu-2\mu}{(2\nu-4\mu)!\,_2F_1(-2k+2\mu,-\nu+3q+2\mu;-2\nu+4\mu;2)\over
\sqrt{(2k-2\mu)!(2\nu-2k-2\mu)!(\nu-3q-2\mu)!(\nu+3q-2\mu)!}}.
\end{eqnarray}
Hence, index of the Kravchuk polynomials serves as additional
quantum number for the SU(3)-degenerated states. In Ref.
\cite{Lashko_NPA} we have shown that a family of eigenfunctions
 of the antisymmetrization operator
combining the states $\Psi_{(2\nu-4\mu,2\mu)_{k}}$ with the same
index $k$ corresponds to a certain binary decay channel of a
three-cluster system into a two-cluster subsystem occurring in a
ground or an excited harmonic-oscillator state and a remaining
cluster. In turn, eigenvalues $\Lambda_{(2\nu-4\mu,2\mu)_{k}}$ of
three-cluster system tend to the eigenvalues of a two-cluster
subsystem with increasing the number of oscillator quanta $\nu$.
However, all eigenvalues of the Pauli-allowed states of $^2n+^2n$
subsystem are equal to unity, in contrast to the eigenvalues of
the $^3$H$+n$ subsystem of the $^5$H considered in
\cite{Lashko_NPA}. By this reason, eigenvalues
$\Lambda_{(2\nu-4\mu,2\mu)_{k}}$ of the $3d'$ system are equal and
tend to unity for $1\leq k\leq \left[\nu-2\mu\over3\right].$
Hence, strictly speaking, the Pauli-allowed states of the $3d'$
system are not uniquely determined and any unitary transformation
can be applied to these states without violating the Pauli
principle. It would seem that a basis of hyperspherical harmonics
also could be used for study of the $3d'$ system. However,
existence of a Pauli-forbidden state makes this statement quite
doubtful. Indeed, since a Pauli-forbidden state and the
Pauli-allowed states have different eigenvalues, no unitary
transformation can leave the diagonal form of the norm kernel
intact. Hence, transformation to the hyperspherical basis could
result in violation of the Pauli exclusion principle.

\section{Norm kernel of three $\alpha$-particles}
\label{sec:6}

Translation-invariant norm kernel for $3\alpha$ system can be
obtained by squaring the norm kernel for three-dineutron system:
\begin{eqnarray*}
I_{3\alpha}\left({\bf A},{\bf B};\tilde{\bf A},\tilde{\bf
B}\right)&=&3!\left[I_{3d'}\left({{\bf A}\over\sqrt{2}}\,,{{\bf
B}\over\sqrt{2}}\,;{\tilde{\bf A}\over\sqrt{2}}\,,{\tilde{\bf
B}\over\sqrt{2}}\right)\right]^2\equiv\sum_{i=1}^6{1\over3!}\hat{{\cal
A}}_i\exp\left\{({\bf A}\tilde{\bf A})+({\bf B}\tilde{\bf
B})\right\}
\end{eqnarray*}
For convenience we divided the antisymmetrization operator
$\hat{{\cal A}}$ of the $3\alpha$-system into six parts
$\hat{{\cal A}}_i$, each corresponding to a certain group of
nucleon permutations\footnote{We denoted an antisymmetrization
operator of $3\alpha$ system by calligraphic  $\hat{{\cal A}}$ in
order to distinguish it from antisymmetrization operator $\hat{A}$
of $3d'$ system}. Six parts $I_{3\alpha}^{(i)}\left({\bf A},{\bf
B};\tilde{\bf A},\tilde{\bf B}\right) $ of the norm kernel
$I_{3\alpha}\left({\bf A},{\bf B};\tilde{\bf A},\tilde{\bf
B}\right)$, which correspond to constituents $\hat{{\cal A}}_i$ of
the antisymmetrization operator $\hat{{\cal A}}$, are given in an
explicit form in Appendix \ref{app:1}.

In this section we are interested in the states of positive parity
and zero orbital momentum. Hence, all the Pauli-allowed states
belong to the $(2\nu-4\mu,2\mu)$ representation of the SU(3)
group, like in the case of $3d'$ system. So specifying operators
$\hat{{\cal A}}_i$ we have taken into account the symmetry
peculiar to the basis functions
$\psi^{\nu-3q-2\mu}_{(2\nu-4\mu,2\mu)}({\bf A},{\bf B})$ on which
these operators should act. Below we shall explain this statement
in detail.

Example of the $3d'$ system illustrates that a permutation of
nucleons belonging to different clusters can be reduced to the
following mathematical operation:
\begin{eqnarray}
f\left({\bf A},{\bf B}\right)\rightarrow const\cdot
f\left(\alpha{\bf A}+\beta{\bf B},\beta^*{\bf A}+\alpha^*{\bf
B}\right), \label{perm_arb}
\end{eqnarray}
where $\alpha,\beta$ are some complex numbers. Hence, due to the
above-mentioned permutation, some rearrangement of the clusters
occurs. Then cyclic permutations of such rearranged clusters can
be described as
\begin{eqnarray*}
f\left({\bf A},{\bf B}\right)\rightarrow const\cdot f\left(e^{\pm
i{2\pi\over3}}(\alpha{\bf A}+\beta{\bf B}),e^{\mp
i{2\pi\over3}}(\beta^*{\bf A}+\alpha^*{\bf B})\right).
\end{eqnarray*}
Obviously, the latter cyclic permutations should give the same
result as the initial permutation (\ref{perm_arb}):
\begin{eqnarray*}
\psi^{\nu-3q-2\mu}_{(2\nu-4\mu,2\mu)}\left(\alpha{\bf A}+\beta{\bf
B},\beta^*{\bf A}+\alpha^*{\bf
B}\right)+\psi^{\nu-3q-2\mu}_{(2\nu-4\mu,2\mu)}\left(e^{i{2\pi\over3}}(\alpha{\bf
A}+\beta{\bf B}),e^{-i{2\pi\over3}}(\beta^*{\bf A}+\alpha^*{\bf
B})\right)+\\
+\psi^{\nu-3q-2\mu}_{(2\nu-4\mu,2\mu)}\left(e^{-i{2\pi\over3}}(\alpha{\bf
A}+\beta{\bf B}),e^{i{2\pi\over3}}(\beta^*{\bf A}+\alpha^*{\bf
B})\right)=3\psi^{\nu-3q-2\mu}_{(2\nu-4\mu,2\mu)}\left(\alpha{\bf
A}+\beta{\bf B},\beta^*{\bf A}+\alpha^*{\bf B}\right).
\end{eqnarray*}
In addition, functions $\psi^{\nu-3q-2\mu}_{(2\nu-4\mu,2\mu)}({\bf
A},{\bf B})$ are symmetric with respect to the interchange of
vectors ${\bf A}$ and ${\bf B}.$\footnote{Such interchange
corresponds to the permutation of two boson clusters.} By this
reason, the following equality holds true:
\begin{eqnarray*}
\psi^{\nu-3q-2\mu}_{(2\nu-4\mu,2\mu)}\left(\beta^*{\bf
A}+\alpha^*{\bf B},\alpha{\bf A}+\beta{\bf
B}\right)=\psi^{\nu-3q-2\mu}_{(2\nu-4\mu,2\mu)}\left(\alpha{\bf
A}+\beta{\bf B},\beta^*{\bf A}+\alpha^*{\bf B}\right).
\end{eqnarray*}
Keeping in mind all of the preceding, let us determine operators
${\hat {\cal A}}_i.$ The first part ${\hat {\cal A}}_1$ of the
antisymmetrization operator ${\hat {\cal A}}$ for three
$\alpha$-particles is very similar to the antisymmetrization
operator ${\hat A}$ of three-dineutron system:
\begin{eqnarray*}
\hat{{\cal A}}_1=6\hat{{\cal A}}_{11}+6\hat{{\cal
A}}_{12}+3\left(\hat{{\cal A}}_{13}^0+\hat{{\cal
A}}_{13}^++\hat{{\cal A}}_{13}^-\right),
\end{eqnarray*}
\begin{eqnarray*}
\hat{{\cal A}}_{11}f({\bf A},{\bf B})=f({\bf A},{\bf
B}),~~\hat{{\cal A}}_{12}f({\bf A},{\bf B})=2f\left(-{{\bf
A}\over2},-{{\bf B}\over2}\right),~~ \hat{{\cal A}}_{13}^0f({\bf
A},{\bf B})=2f\left({{\bf A}+{\bf B}\over2},{{\bf A}+{\bf
B}\over2}\right).
\end{eqnarray*}
Operators $\hat{A}_{ij}^\pm$ are defined as follows:
\begin{eqnarray*}
\hat{{\cal A}}_{ij}^\pm f({\bf A},{\bf
B})=\left[\hat{A}_{ij}^0f({\bf A},{\bf B})\right]_{{\bf
A}\rightarrow\exp\left(\pm\,i{2\pi\over 3}\right){\bf A},{\bf
B}\rightarrow\exp\left(\mp i{2\pi\over 3}\right){\bf B}}.
\end{eqnarray*}
Permutations belonging to this group differ from the permutations
in three-dineutron system only in one respect: the latter refer to
fermions, while the former relate to bosons. Considering
dineutrons and diprotons as bosons, we come to the conclusion that
$\hat{A}_{11}$ corresponds to the permutations of
$\alpha$-clusters, ${\hat A}_{12}$ can be matched with the
exchange of dineutrons and diprotons between all
$\alpha$-particles, while ${\hat A}_{13}^0$ and ${\hat
A}_{13}^\pm$ are associated with the permutations of dineutrons
and diprotons between two $\alpha$-clusters with the third cluster
being a spectator.

The second part $\hat{{\cal A}}_2$ of the antisymmetrization
operator of $3\alpha$ system
\begin{eqnarray*}
\hat{{\cal A}}_2 =6\hat{{\cal A}}_{21}+6\hat{{\cal
A}}_{22}+3(\hat{{\cal A}}_{23}^0+\hat{{\cal A}}_{23}^++\hat{{\cal
A}}_{23}^-)
\end{eqnarray*}
combines permutations involving deuteron clusters:
\begin{eqnarray*}
\hat{{\cal A}}_{21} f({\bf A},{\bf B})=4f\left(-{{\bf
A}\over2},-{{\bf B}\over2}\right),~\hat{{\cal A}}_{22}f({\bf
A},{\bf B})=8f\left({{\bf A}\over4},{{\bf B}\over4}\right),~~
\hat{{\cal A}}_{23}^0f({\bf A},{\bf B})=8f\left(-{{\bf A}+{\bf
B}\over4},-{{\bf A}+{\bf B}\over4}\right);
\end{eqnarray*}
$\hat{{\cal A}}_{21}$ includes permutations of deuteron clusters
with unit spin projection; $\hat{{\cal A}}_{22}$ is associated
with the exchange of nucleons between three deuterons, while the
remaining three deuterons (characterized by zero spin projection)
are spectators; $\hat{{\cal A}}_{23}^0,\hat{{\cal A}}_{23}^\pm$
combine permutations resulting in the destruction of one
$\alpha$-cluster with the subsequent distribution of two
constituent deuterons with unit spin projection between two other
$\alpha$-clusters and formation of the cluster from two remaining
deuterons with zero spin projections.

Operator $\hat{{\cal A}}_{3}$
\begin{eqnarray*}
\hat{{\cal A}}_3 =6\left(\hat{{\cal A}}_{31}+\hat{{\cal
A}}_{32}+\hat{{\cal
A}}_{33}+\hat{{\cal A}}_{34}^0+\hat{{\cal A}}_{34}^++\hat{{\cal
A}}_{34}^-\right)
\end{eqnarray*}
includes permutations of protons, with dineutrons remaining
intact:
\begin{eqnarray*}
\hat{{\cal A}}_{31}f({\bf A},{\bf B})=4f\left({{\bf
A}\over4},{{\bf B}\over4}\right);~~ \hat{{\cal A}}_{32}f({\bf
A},{\bf B})=4f\left({5+i\sqrt{3}\over8}{\bf
A},{5-i\sqrt{3}\over8}{\bf B}\right);\\
\hat{{\cal A}}_{33}f({\bf A},{\bf
B})=4f\left({5-i\sqrt{3}\over8}{\bf A},{5+i\sqrt{3}\over8}{\bf
B}\right);~~ \hat{{\cal A}}_{34}^0f({\bf A},{\bf
B})=4f\left({2{\bf A}-{\bf B}\over4},-{{\bf A}-2{\bf
B}\over4}\right).
\end{eqnarray*}
$\hat{{\cal A}}_{31}$ combines simultaneous cyclic permutations of
spin-up and spin-down protons; $\hat{{\cal A}}_{32}$ and
$\hat{{\cal A}}_{33}$ contain cyclic permutations of spin-down or
spin-up protons, accordingly; $\hat{{\cal A}}_{34}^0,\hat{{\cal
A}}_{34}^\pm$ are responsible for the one-proton exchange between
one cluster and each of the remaining two clusters.

The next part of antisymmetrization operator $\hat{{\cal A}}$ is
related to the exchange of spin-down protons between two clusters
accompanied by the identity or cyclic permutations of dineutrons.
\begin{eqnarray*}
\hat{{\cal A}}_{4}=6(\hat{{\cal A}}_{41}^0+\hat{{\cal A}}_{41}^++\hat{{\cal A}}_{41}^-+\hat{{\cal A}}_{42}^0+\hat{{\cal A}}_{42}^++\hat{{\cal A}}_{42}^-+\hat{{\cal A}}_{43}^0+\hat{{\cal A}}_{43}^++\hat{{\cal A}}_{41}^-)
\end{eqnarray*}
In this case  dineutrons and one of diprotons are not rearranged.
\begin{eqnarray*}
\hat{{\cal A}}_{41}^0f({\bf A},{\bf B})=-4f\left({3{\bf A}+{\bf
B}\over4},{{\bf A}+3{\bf B}\over4}\right),\\
\hat{{\cal A}}_{42}^0f({\bf A},{\bf B})=-4f\left({i\sqrt{3}{\bf A}+{\bf
B}\over4},{{\bf A}-i\sqrt{3}{\bf B}\over4}\right),~~
\hat{{\cal A}}_{43}^0f({\bf A},{\bf B})=-4f\left({-i\sqrt{3}{\bf A}+{\bf
B}\over4},{{\bf A}+i\sqrt{3}{\bf B}\over4}\right).
\end{eqnarray*}
$\hat{{\cal A}}_{41}$, $\hat{{\cal A}}_{42}$ and $\hat{{\cal
A}}_{43}$ are associated with the identity, clockwise or
contraclockwise cyclic permutations of dineutrons, accordingly.

Operator $\hat{{\cal A}}_{5}$ is characterized by the odd number
of pair permutations of nucleons:
\begin{eqnarray*}
\hat{{\cal A}}_{5}=6(3\hat{{\cal A}}_{51}+\hat{{\cal
A}}_{52}^0+\hat{{\cal A}}_{52}^++\hat{{\cal A}}_{52}^-+\hat{{\cal
A}}_{53}^0+\hat{{\cal A}}_{53}^++\hat{{\cal A}}_{53}^-).
\end{eqnarray*}
In this case one diproton remains a spectator.
\begin{eqnarray*}
\hat{{\cal A}}_{51}f({\bf A},{\bf B})=-8f\left({{\bf
A}\over4},{{\bf B}\over4}\right),~~ \hat{{\cal A}}_{52}^0f({\bf
A},{\bf B})=-8f\left({1\over4}{\bf
A}+{i\sqrt{3}\over4}e^{-i2\pi/3}{\bf
B},-{i\sqrt{3}\over4}e^{i2\pi/3}{\bf A}+{1\over4}{\bf B}\right),\\
\hat{{\cal A}}_{53}^0f({\bf A},{\bf B})=-8f\left({1\over4}{\bf
A}-{i\sqrt{3}\over4}e^{i2\pi/3}{\bf
B},{i\sqrt{3}\over4}e^{-i2\pi/3}{\bf A}+{1\over4}{\bf B}\right).
\end{eqnarray*}
$\hat{{\cal A}}_{51}$ combines permutations resulting in the
two-nucleon exchange between all clusters; $\hat{{\cal
A}}_{52}^0,\hat{{\cal A}}_{52}^\pm$  and $\hat{{\cal
A}}_{53}^0,\hat{{\cal A}}_{53}^\pm$ correspond to the one-nucleon
exchange between two clusters accompanied by the exchange of
deuterons with unit or zero spin projection, accordingly.

The last part $\hat{{\cal A}}_{6}$ of the antisymmetrization
operator looks like
\begin{eqnarray*}
\hat{{\cal A}}_{6}=3(\hat{{\cal A}}_{61}^0+\hat{{\cal
A}}_{61}^++\hat{{\cal A}}_{61}^-+\hat{{\cal A}}_{62}^0+\hat{{\cal
A}}_{62}^++\hat{{\cal A}}_{62}^-)+6(\hat{{\cal
A}}_{63}^0+\hat{{\cal A}}_{63}^++\hat{{\cal A}}_{63}^-).
\end{eqnarray*}
All constituent parts of operator $\hat{{\cal A}}_{6}$ can be
expressed via previously defined permutation operators:
\begin{eqnarray*}
\hat{{\cal A}}_{61}^0=2\hat{{\cal A}}_{13}^0,~~ \hat{{\cal
A}}_{62}^0=2\hat{{\cal A}}_{23}^0,~~\hat{{\cal
A}}_{63}^0=2\hat{{\cal A}}_{34}^0.
\end{eqnarray*}
Taking into account that $\hat{{\cal A}}_{21}=2\hat{{\cal
A}}_{12}$ and $\hat{{\cal A}}_{22}=2\hat{{\cal A}}_{31},$ we come
to the final expression for the antisymmetrization operator of the
$3\alpha$ system:
\begin{eqnarray*}
\hat{{\cal A}}&=&\sum_{i=1}^6{1\over3!}\hat{{\cal A}}_i=\hat{{\cal
A}}_{11}+3\hat{{\cal A}}_{12}+ 3\hat{{\cal A}}_{31}+\hat{{\cal
A}}_{32}+\hat{{\cal
A}}_{33}+3\hat{{\cal A}}_{51}+\\
&+&{3\over2}(\hat{{\cal A}}_{13}^0+\hat{{\cal
A}}_{13}^++\hat{{\cal A}}_{13}^-)+ {3\over2}(\hat{{\cal
A}}_{23}^0+\hat{{\cal A}}_{23}^++\hat{{\cal
A}}_{23}^-)+3(\hat{{\cal
A}}_{34}^0+\hat{{\cal A}}_{34}^++\hat{{\cal A}}_{34}^-)+\\
&+&\hat{{\cal A}}_{41}^0+\hat{{\cal A}}_{41}^++\hat{{\cal
A}}_{41}^-+\hat{{\cal A}}_{42}^0+\hat{{\cal A}}_{42}^++\hat{{\cal
A}}_{42}^-+\hat{{\cal A}}_{43}^0+\hat{{\cal A}}_{43}^++\hat{{\cal
A}}_{43}^-+ \hat{{\cal A}}_{52}^0+\hat{{\cal A}}_{52}^++\hat{{\cal
A}}_{52}^-+\hat{{\cal A}}_{53}^0+\hat{{\cal A}}_{53}^++\hat{{\cal
A}}_{53}^-.
\end{eqnarray*}
Applying operator $\hat{{\cal A}}$ to the basis functions
$\psi^{\nu-3q-2\mu}_{(2\nu-4\mu,2\mu)}({\bf A},{\bf B}),$ we shall
come to the following expansion of the partial norm kernel
$i^{3\alpha}_{(2\nu-4\mu,2\mu)}$ for the $3\alpha$-system:
\begin{eqnarray}
i^{3\alpha}_{(2\nu-4\mu,2\mu)}&=&\sum_{q=0}^{[{\nu-2\mu\over3}]}\sum_{\tilde{q}=0}^{[{\nu-2\mu\over3}]-q}
\langle2\nu,\nu-3q-2\mu|2\nu,\nu-3q-3\tilde{q}-2\mu\rangle_{3\alpha}\times\nonumber\\
& &\times
\left(\psi_{(2\nu-4\mu,2\mu)}^{\nu-3q-2\mu}\tilde{\psi}_{(2\nu-4\mu,2\mu)}^{\nu-3q-3\tilde{q}-2\mu}+
\psi_{(2\nu-4\mu,2\mu)}^{\nu-3q-3\tilde{q}-2\mu}\tilde{\psi}_{(2\nu-4\mu,2\mu)}^{\nu-3q-2\mu}\right).
\label{part_norm_12C}
\end{eqnarray}
The above expansion differs from similar expression
(\ref{part_norm_6n}) for the three-dineutron system only in the
explicit form of the matrix elements of the secular equation:
\begin{eqnarray*}
\langle2\nu,\nu-3q-2\mu|2\nu,\nu-3q-3\tilde{q}-2\mu\rangle_{3\alpha}=\\
=\left(1+{6\over4^\nu}\left(1-{2\over4^\nu}\right)+
{1\over8^{6q-1}}\left({7\over16}\right)^{\nu-3q}\sum_{m=0}^{3q}
\left(
\begin{array}{c}
  6q \\
  2m \\
\end{array}\right)
(-1)^m5^{6q-2m}3^m\right)\delta_{\tilde{q},0}+\\
+{9\over4^{2\nu-1}}\sqrt{{(\nu+3q+3\tilde{q}-2\mu)!(\nu-3q-3\tilde{q}-2\mu)!\over{(\nu+3q-2\mu)!(\nu-3q-2\mu)!}(1+\delta_{q,0})
(1+\delta_{\tilde{q},-q})}}\,
\left(
\begin{array}{c}
  \nu+3q-2\mu \\
  \nu-3q-3\tilde{q}-2\mu \\
\end{array}\right)
\times\\
\times\left\{3^{2\mu}(-1)^{\tilde{q}}\left[2^{6q+3\tilde{q}}
{_2}F_1(\alpha,\beta;\gamma;4)+ 2^{2\nu-6q-3\tilde{q}-4\mu}
{_2}F_1\left(\alpha,\beta;\gamma;{1\over4}\right)\right]\right.-\\
-{1\over3}8^{2\mu}\left[3^{6q+3\tilde{q}}
{_2}F_1(\alpha,\beta;\gamma;9)+ 3^{2\nu-6q-3\tilde{q}-4\mu}
{_2}F_1\left(\alpha,\beta;\gamma;{1\over9}\right)\right]-\\
\left.
-\left.2^{2\mu}\left(1+(-1)^{\tilde{q}}\right)\left[(i\sqrt{3})^{6q+3\tilde{q}}
{_2}F_1(\alpha,\beta;\gamma;3)+
3^{\nu-2\mu}(-1)^q(i\sqrt{3})^{-6q-3\tilde{q}}{_2}F_1\left(\alpha,\beta;\gamma;{1\over3}\right)\right]\right\}\right),
\end{eqnarray*}
where indices of hypergeometric function are defined as follows:
\begin{eqnarray*}
\alpha=-\nu+3q+2\mu,~~\beta=-\nu+3q+3\tilde{q}+2\mu,~~\gamma=6q+3\tilde{q}+1.
\end{eqnarray*}
Eigenfunctions $\Psi_{(2\nu-4\mu,2\mu)}$ of the antisymmetrization
operator $\hat{{\cal A}}$ for the $3\alpha$-system are given in
terms of orthogonal polynomials of a discrete variable, as in the
case of the $3d'$-system (see Eq. (\ref{wf_expand})). The number
of different states possessing SU(3) symmetry $(2\nu-4\mu,2\mu)$,
i.e., the degree of the SU(3)-degeneracy, is the same for
three-boson system, $3d'$-system and $3\alpha$-system and equal to
$[{\nu-2\mu\over3}]+1.$ However, some of these eigenfunctions
correspond to zero eigenvalues and thus they are forbidden by the
Pauli principle. The number of the latter functions depends on the
system considered and increases with increasing the number of
nucleons in each cluster. For example, in the three-boson system
all the states belonging to the $(2\nu-4\mu,2\mu)$ representation
have been allowed by the Pauli principle, while in the
three-dineutron system at a given number of oscillator quanta
$\nu$ there is a Pauli-forbidden state $\Psi_{(2\nu,0)_{k=0}}.$ In
the $3\alpha$-system at $\nu\geq3$ there are two Pauli-forbidden
states belonging to the $(2\nu,0)$ SU(3)-representation and one
Pauli-forbidden state characterized by the $(2\nu-4,2)$
SU(3)-indices.
\begin{table*}[htb]
\caption{\label{table:1} Eigenvalues
$\Lambda_{(2\nu-4\mu,2\mu)_{k}}$ belonging to the first three
families of the Pauli-allowed states of the $3\alpha$-system}
\begin{tabular}{|c|ccc|cccc|ccccc|}
\hline

 & \multicolumn{3}{c}{$2k=4$} & \multicolumn{4}{c}{$2k=6$} & \multicolumn{5}{c}{$2k=8$} \\

$\nu$ & $\mu=2$ & $\mu=1$ & $\mu=0$ & $\mu=3$ & $\mu=2$ & $\mu=1$
& $\mu=0$ & $\mu=4$ & $\mu=3$ & $\mu=2$ &
$\mu=1$ & $\mu=0$ \\

\hline
4 & 0.5933 &  &  &  &  &  &  &  &  &  &  & \\

5 & 0.6748 & 0.4894 &  &  &  &  &  &  &  &  &  & \\

6 & 0.7161 & 0.5647 & 0.6497 & 0.8714 &  &  &  &  &  &  &  &  \\

7 & 0.7366 & 0.6333 & 0.6269 & 0.9081 & 0.8657 &  &  &  &  &  &  &  \\

8 & 0.7450 & 0.6859 & 0.6457 & 0.9246 & 0.8852 & 0.9073 &  & 0.9640 &  &  &  & \\

9 & 0.7482 & 0.7196 & 0.6773 & 0.9324 & 0.9042 & 0.8959 & 0.9478 & 0.9754 & 0.9666 &  &  & \\

10 & 0.7493 & 0.7372 & 0.7071 & 0.9356 & 0.9191 & 0.9000 & 0.9193 & 0.9805 & 0.9712 & 0.9770 &  & \\

11 & 0.7497 & 0.7449 & 0.7283 & 0.9368 & 0.9287 & 0.9106 & 0.9084 & 0.9828 & 0.9759 & 0.9738 & 0.9865 & \\

12 & 0.7499 & 0.7480 & 0.7404 & 0.9372 & 0.9338 & 0.9212 & 0.9093 & 0.9838 & 0.9796 & 0.9747 & 0.9792 & 0.9906\\

13 & 0.7500 & 0.7492 & 0.7461 & 0.9374 & 0.9360 & 0.9290 & 0.9159 & 0.9842 & 0.9821 & 0.9774 & 0.9763 & 0.9837\\

14 & 0.7500 & 0.7497 & 0.7485 & 0.9375 & 0.9369 & 0.9336 & 0.9236 & 0.9843 & 0.9834 & 0.9801 & 0.9764 & 0.9792\\

15 & 0.7500 & 0.7499 & 0.7494 & 0.9375 & 0.9373 & 0.9358 & 0.9298 & 0.9843 & 0.9840 & 0.9821 & 0.9783 & 0.9773\\

16 & 0.7500 & 0.7499 & 0.7498 & 0.9375 & 0.9374 & 0.9368 & 0.9337 & 0.9844 & 0.9842 & 0.9833 & 0.9804 & 0.9775\\

17 & 0.7500 & 0.7500 & 0.7499 & 0.9375 & 0.9375 & 0.9372 & 0.9358 & 0.9844 & 0.9843 & 0.9839 & 0.9822 & 0.9790\\

18 & 0.7500 & 0.7500 & 0.7500 & 0.9375 & 0.9375 & 0.9374 & 0.9368 & 0.9844 & 0.9844 & 0.9842 & 0.9833 & 0.9808\\

19 & 0.7500 & 0.7500 & 0.7500 & 0.9375 & 0.9375 & 0.9375 & 0.9372 & 0.9844 & 0.9844 & 0.9843 & 0.9839 & 0.9823\\

20 & 0.7500 & 0.7500 & 0.7500 & 0.9375 & 0.9375 & 0.9375 & 0.9374 & 0.9844 & 0.9844 & 0.9843 & 0.9842 & 0.9833\\

\hline
\end{tabular}
\end{table*}
All the Pauli-allowed states of the $3\alpha$-system have distinct
eigenvalues, which tend to the eigenvalues of a $2\alpha$
subsystem with increasing the number of oscillator quanta $\nu$
(see Table \ref{table:1}):
\begin{eqnarray*}
\lim_{\nu-2\mu\to\infty}\Lambda^{{^{12}}{\rm
C}=3\alpha}_{(2\nu-4\mu,2\mu)_{k}}\rightarrow\lambda_{2k}^{{^{8}}{\rm
Be}=2\alpha}=1-\left({1\over4}\right)^{k-1}+3\delta_{k,0}.
\end{eqnarray*}

Eigenfunctions $\Psi_{(2\nu-4\mu,2\mu)_k}$ also become more
complicated with increasing the number of nucleons. In the case of
$3\alpha$-system each eigenfunction of the antisymmetrization
operator is a superposition of functions
$\psi^{\nu-3q-2\mu}_{(2\nu-4\mu,2\mu)}$ with all nonvanishing
expansion coefficients $D^{\nu-3q-2\mu}_{(2\nu-4\mu,2\mu)_{k}}.$
However, as the eigenvalues $\Lambda_{(2\nu-4\mu,2\mu)_{k}}$ of
the $3\alpha$ norm kernel approach limit values
$\lambda_{2k}^{{^{8}}{\rm Be}}$, corresponding eigenvectors
$\Psi_{(2\nu-4\mu,2\mu)_{k}}$ take simple analytical form
$\Psi_{(2\nu-4\mu,2\mu)_{k}}^{\rm as}$ (see Eq. (\ref{func_as})),
as can be seen from Table \ref{table:2}. Consequently, in the
limit $\nu\gg k$ the expansion coefficients
$D^{\nu-3q-2\mu}_{(2\nu-4\mu,2\mu)_{k}}$ can be identified with
the Kravchuk polynomials of a discrete variable:
\begin{eqnarray*}
D^{\nu-3q-2\mu}_{(2\nu-4\mu,2\mu)_{k}}\rightarrow\sqrt{6\over1+\delta_{q,0}}\,\bar{{\cal
K}}^{1/2}_{2k}(\nu-3q-2\mu),
\end{eqnarray*}
where the explicit form of the Kravchuk polynomials $\bar{{\cal
K}}^{1/2}_{2k}(\nu-3q-2\mu)$ is given by Eq. (\ref{Kravchuk}).

Therefore, index $k,$  which makes sense of the number of
oscillator quanta accounted for by the $^8$Be subsystem, serves as
an additional quantum number for the SU(3)-degenerated states. It
should be noted also that due to the incoincidence of eigenvalues
$\Lambda^{{^{12}}{\rm C}=3\alpha}_{(2\nu-4\mu,2\mu)_{k}},$ the
eigenfunctions $\Psi_{(2\nu-4\mu,2\mu)_k}$ are uniquely
determined. Any unitary transformation applied to the latter
SU(3)-basis functions would disrupt the diagonal form of the norm
kernel of the $3\alpha$ system and, hence, is inappropriate in
this case.

Noteworthy also is a remarkably simple form of the eigenfunctions
$\Psi_{(2\nu-4\mu,2\mu)_{k=\mu}}$ in the case $\nu-2\mu<3:$
\begin{eqnarray*}
\Psi_{(2\nu-4\mu,2\mu)_{k=\mu}}=\psi^{\nu-2\mu}_{(2\nu-4\mu,2\mu)},~~\left[{\nu-2\mu\over3}\right]=0.
\end{eqnarray*}
It is precisely this fact that explains why the overlap integrals
$\int\Psi_{(2\nu-4\mu,2\mu)_{k=\mu}}\Psi_{(2\nu-4\mu,2\mu)_{k=\mu}}^{\rm
as}d\mu_B$ are equal to unity, as long as $\nu-2\mu<3.$

In turn, the eigenvalues $\Lambda_{(2\nu-4\mu,2\mu)_{k=\mu}}$ are
given by the following analytical expression, provided that
$\nu-2\mu<3:$
\begin{eqnarray*}
\Lambda_{(2\nu-4\mu,2\mu)_{k=\mu}}&=&1+{6\over4^\nu}\left(1-{2\over4^\nu}\right)+
8\left({7\over16}\right)^{\nu}+{18\over4^{2\nu}}\times\\
&\times&\left\{3^{2\mu}\left[{_2}F_1(-\nu+2\mu,-\nu+2\mu;1;4)+
4^{\nu-2\mu}
{_2}F_1\left(-\nu+2\mu,-\nu+2\mu;1;{1\over4}\right)\right] -\right.\\
&-& \left.
{1\over3}8^{2\mu}\left[{_2}F_1(-\nu+2\mu,-\nu+2\mu;1;9)+
9^{\nu-2\mu}
{_2}F_1\left(-\nu+2\mu,-\nu+2\mu;1;{1\over9}\right)\right]-\right.\\
&-&\left.2^{2\mu+1} \left[{_2}F_1(-\nu+2\mu,-\nu+2\mu;1;3)+
3^{\nu-2\mu}{_2}F_1\left(-\nu+2\mu,-\nu+2\mu;1;{1\over3}\right)\right]\right\}
\end{eqnarray*}

\begin{table*}[htb]
\caption{\label{table:2} Overlap integrals
$\int\Psi_{(2\nu-4\mu,2\mu)_{k}}\Psi_{(2\nu-4\mu,2\mu)_{k}}^{\rm
as}d\mu_B$ versus the number of quanta $\nu$ for the first three
families of the Pauli-allowed states of the $3\alpha$-system}
\begin{tabular}{|c|ccc|cccc|ccccc|}
\hline

 & \multicolumn{3}{c}{$2k=4$} & \multicolumn{4}{c}{$2k=6$} & \multicolumn{5}{c}{$2k=8$} \\

$\nu$ & $\mu=2$ & $\mu=1$ & $\mu=0$ & $\mu=3$ & $\mu=2$ & $\mu=1$
& $\mu=0$ & $\mu=4$ & $\mu=3$ & $\mu=2$ &
$\mu=1$ & $\mu=0$ \\

\hline
4 & 1.0000 &  &  &  &  &  &  &  &  &  &  & \\

5 & 1.0000 & 0.9670 &  &  &  &  &  &  &  &  &  & \\

6 & 1.0000 & 0.9957 & 0.8688 & 1.0000 &  &  &  &  &  &  &  &  \\

7 & 0.9988 & 0.9994 & 0.9661 & 1.0000 & 0.9535 &  &  &  &  &  &  &  \\

8 & 0.9996 & 0.9944 & 0.9919 & 1.0000 & 0.9929 & 0.8127 &  & 1.0000 &  &  &  & \\

9 & 0.9999 & 0.9940 & 0.9960 & 0.9977 & 0.9990 & 0.9321 & 0.7255 & 1.0000 & 0.9474 &  &  & \\

10 & 1.0000 & 0.9973 & 0.9949 & 0.9993 & 0.9932 & 0.9800 & 0.8677 & 1.0000 & 0.9914 & 0.8064 &  & \\

11 & 1.0000 & 0.9993 & 0.9958 & 0.9998 & 0.9923 & 0.9920 & 0.9421 & 0.9969 & 0.9987 & 0.9269 & 0.7151 & \\

12 & 1.0000 & 0.9998 & 0.9981 & 1.0000 & 0.9965 & 0.9912 & 0.9790 & 0.9989 & 0.9928 & 0.9778 & 0.8537 & 0.6447\\

13 & 1.0000 & 1.0000 & 0.9994 & 1.0000 & 0.9990 & 0.9924 & 0.9908 & 0.9998 & 0.9920 & 0.9913 & 0.9311 & 0.8009\\

14 & 1.0000 & 1.0000 & 0.9998 & 1.0000 & 0.9998 & 0.9962 & 0.9925 & 1.0000 & 0.9963 & 0.9907 & 0.9736 & 0.8826\\

15 & 1.0000 & 1.0000 & 1.0000 & 1.0000 & 1.0000 & 0.9988 & 0.9942 & 1.0000 & 0.9990 & 0.9919 & 0.9885 & 0.9399\\

16 & 1.0000 & 1.0000 & 1.0000 & 1.0000 & 1.0000 & 0.9997 & 0.9970 & 1.0000 & 0.9998 & 0.9959 & 0.9909 & 0.9741\\

17 & 1.0000 & 1.0000 & 1.0000 & 1.0000 & 1.0000 & 0.9999 & 0.9988 & 1.0000 & 0.9999 & 0.9987 & 0.9928 & 0.9878\\

18 & 1.0000 & 1.0000 & 1.0000 & 1.0000 & 1.0000 & 1.0000 & 0.9997 & 1.0000 & 1.0000 & 0.9996 & 0.9962 & 0.9914\\

19 & 1.0000 & 1.0000 & 1.0000 & 1.0000 & 1.0000 & 1.0000 & 0.9999 & 1.0000 & 1.0000 & 0.9999 & 0.9986 & 0.9937\\

20 & 1.0000 & 1.0000 & 1.0000 & 1.0000 & 1.0000 & 1.0000 & 1.0000 & 1.0000 & 1.0000 & 1.0000 & 0.9996 & 0.9966\\

\hline
\end{tabular}
\end{table*}

From Table \ref{table:1} and Table \ref{table:2} we notice that
the Pauli-allowed basis states $\Psi_{(2\nu-4\mu,2\mu)_{k}}$ can
be arranged into the branches and families, with all the states of
a particular branch having common SU(3)-symmetry index $\mu$ and
overlapping generously with corresponding asymptotic function
(\ref{func_as}), but differing in value of the first index
$\bar{\lambda}$ of the SU(3)-symmetry. The eigenvalues belonging
to a given branch tend to the same limit value
$\lambda_{2k}^{{^{8}}{\rm Be}}$ with the number of quanta
increasing. The branches which share limit eigenvalues are
combined in the family of the eigenstates, which thus is
completely determined by the degree $k$ of the corresponding
Kravchuk polynomial.

It should be noted that the states listed in Table \ref{table:1}
and Table \ref{table:2} exhaust all possible basis functions
allowed by the Pauli principle, as long as $\nu\leq9$. As $\nu$
increases, new families of states characterized by $2k>8$ also
appear. However, eigenstates and eigenvalues belonging to such
families are governed by the same laws as those tabulated in
Tables \ref{table:1}, \ref{table:2}. The state
$\Psi_{{(0,4)}_{k=2}}$ characterized by the minimum number of
quanta $\nu=4$ and belonging to the $\mu=2$ branch of the
$k=k_{\rm min}=2$ family of the Pauli-allowed states is identical
to the state of the leading SU(3)-representation of the oscillator
translation-invariant shell model in the Elliott scheme. Other
states belonging to the $\mu=2$ branch provide the main
corrections to the states of the Elliott model resulting from the
$\alpha$-clustering of the $^{12}$C nucleus and its decay through
the $^{12}$C$\rightarrow^8$Be$+\alpha$ channel. Branches of the
states belonging to the families characterized by $k>k_{\rm min}$
reproduce the excitations of the $^8$Be accompanied by the
increasing of the number of oscillator quanta accounted for by
this subsystem.

Existence of different limit eigenvalues $\lambda_{2k}^{{^{8}}{\rm
Be}}$ reflects the possibility for two $\alpha$-clusters to be
close to each other and far apart from the third $\alpha$-cluster.
Owing to this the unitary transformation from the SU(3)-basis to
the other one is rendered possible only within a particular family
of the Pauli-allowed states. Transformation to the three-cluster
hyperspherical harmonics, contrastingly, involves the states
belonging to different families. The hypermomentum can not serve
as quantum number of Pauli-allowed basis functions, because the
Pauli exclusion principle mixes the states with different values
of hypermomentum. Hence, restriction for the maximum value of
hypermomentum $K$ included in calculation $K\leq K_{max}$ leads to
spoiling the Pauli-allowed basis functions corresponding to the
number of oscillator quanta $\nu>K_{max}$, and this effect is
enhanced with increasing $\nu$. The angular-momentum-coupled basis
can be used instead in the region $\nu\gg k$, where all
eigenvalues $\Lambda_{(2\nu-4\mu,2\mu)_{k}}$ belonging to the
$k_{\rm th}$ family coincide with their limit values
$\lambda_{2k}^{{^{8}}{\rm Be}}$. Basis functions of the latter
basis are labeled by the number of quanta, and the angular momenta
of the $^{8}{\rm Be}$ subsystem coinciding with the angular
momentum of the relative motion of this subsystem and a remaining
$\alpha$-particle\footnote{Recall that only the states with total
angular momentum $L=0$ are considered in this paper.}. The
transformation to this basis is performed within a particular
family of the Pauli-allowed states and, therefore, does not result
in violation of the Pauli principle. At the same time, the
angular-momentum-coupled basis allows one to set the asymptotic
boundary conditions for the expansion coefficients of a
three-cluster wave function in the continuum. This point has been
discussed in detail in Ref. \cite{Lashko_NPA} with the $^3$H$+n+n$
system.

From the above discussion it appears that asymptotically each
family of Pauli-allowed states of $3\alpha$-system corresponds to
a certain binary decay channel of the latter system into a
$2\alpha$ subsystem occurring in a ground or an excited
harmonic-oscillator state and a remaining $\alpha$-particle.
Hence, the excited states of the $^{12}$C nucleus should decay via
subsequent stage
$^{12}$C$\rightarrow^8$Be$+\alpha\rightarrow\alpha+\alpha+\alpha$
rather than in a "democratic" way
$^{12}$C$\rightarrow\alpha+\alpha+\alpha$ usually associated with
hyperspherical harmonics. The latter basis can, probably, provide
an adequate description of the ground state of the $^{12}$C
nucleus, but it is definitely not appropriate for studying the
continuum states.

Finally, let us compare the behavior of the eigenvalues and
eigenfunctions of the $3\alpha$ system with those of the
$^5$H$=^3$H$+n+n$ system discussed in detail in Ref.
\cite{Lashko_NPA}. First of all, at a given number of quanta $\nu$
the number of the Pauli-allowed states in the $3\alpha$ system is
much less than that of the $^5$H system. In the $3\alpha$ system
only the families of the states corresponding to even values of
quantum number $k$ are consistent with the requirements of the
Pauli exclusion principle, while the model space of the
Pauli-allowed states of $^3$H$+n+n$ system includes the families
characterized by both odd and even values of $k$. The number of
the Pauli-forbidden states in the $3\alpha$-system is also greater
than in the $^5$H nucleus. In the $^5$H system only one branch
falls into the category of the Pauli-forbidden states. It is
characterized by $k=0$ and corresponds to the occupation of an
$s$-state in the $^4$H subsystem by one of the valence neutrons.
The $3\alpha$ system have three branches of the Pauli-allowed
states: a branch being a member of $k=0$ family and two branches
entering in the family $2k=2.$ These two families correspond to
two forbidden states of the $2\alpha$ subsystem. Moreover, whereas
in the case of the $^5$H nucleus all branches belonging to the
same family appear at the same number of quanta, the branches
combining into a particular family of the $3\alpha$ system start
in one-quantum increment. A new family of the Pauli-allowed states
in the $^5$H system arises with increasing the number of quanta by
one, while families of such states in the $3\alpha$ system emerge
in three-quantum interval. This results in decreasing the model
space of the Pauli-allowed functions in the internal region of
small distance between clusters in the $3\alpha$ system as
compared with the $^5$H system.

All the differences discussed above are due to a high symmetry of
the $3\alpha$-system, which comprises of three identical clusters,
while the $^3$H$+n+n$ system have only two identical constituents.
In addition, in the $3\alpha$-system each family begins with the
branch having maximum possible SU(3)-symmetry index $\mu=k.$ This
fact gives grounds to expect that the branches characterized by
$\mu=k$ should be more favorable than the branches with $\mu=0,$
in contrast to the case of the $^5$H nucleus. Of course, both for
the case of the $3\alpha$ system and the $^3$H$+n+n$ system the
families of the allowed states associated with the low-order
Kravchuk polynomials should dominate.

The eigenvalues of the antisymmetrization operator of the
$3\alpha$ system tend to unity from below resulting in the
repulsive effective interaction between $\alpha$-clusters due to
the change in the kinetic energy of relative motion under the
effect of the Pauli exclusion principle (see Refs.
\cite{Lashko_NPA, Lashko_PRC} for details). Hence taking into
account all the exchange effects in $3\alpha$-system may allows
one to overcome the problem of overbinding of the $^{12}$C nucleus
without introducing an additional three-cluster repulsion.

\section{Conclusion}
\label{sec:7}

Within the microscopic model based on the algebraic version of the
resonating group method the role of the Pauli principle in the
formation of continuum wave function of nuclear systems composed
of three identical $s$-clusters has been investigated. Our
principal concern has been with the study of the exchange effects
contained in the genuine three-cluster norm kernel, i.e., taking
into account the eigenvalues of the Pauli-allowed states.

Norm kernels for the system of three identical fermions and the
system of three identical bosons have been constructed in the
Fock-Bargmann space. It has been shown that the former serves as
the main building block for the norm kernels of nuclear systems
composed of three identical clusters, while the latter generates
the basis functions needed for the description of nuclear systems
composed of three identical boson clusters (such as
$^6n=^2n+^2n+^2n$ and $^{12}$C=$\alpha+\alpha+\alpha$). Simple
analytical method of constructing the norm kernel for $3d'$ system
and $3\alpha$ system in the Fock-Bargmann space has been suggested
and realized.

The Fock-Bargmann image of the antisymmetrization operator has
been found for three-fermion, three-boson, three-dineutron and
$3\alpha$ systems. Classification of different constituents of the
antisymmetrization operator by belonging to a certain group of
permutations is given. Careful analysis of the structure of the
eigenfunctions and behavior of the eigenvalues of the
antisymmetrizer has been performed for all the above-mentioned
three-cluster systems. It has been demonstrated that the
Pauli-allowed states of three-cluster systems composed of
identical boson clusters can be represented as a superposition of
the eigenfunctions of the three-boson norm kernel, with the
expansion coefficients being the orthogonal polynomials of a
discrete variable.

The Pauli-allowed basis functions for the $3\alpha$ and the $3d'$
systems are given in an explicit form and the asymptotic behavior
of these functions is established. Eigenvalues of the
three-cluster system are shown to tend to the eigenvalues of a
two-cluster subsystem with increasing the number of oscillator
quanta $\nu$. At the same time, the corresponding eigenvectors
take simple analytical form, while the expansion coefficients
become the Kravchuk polynomials as the number of oscillator quanta
increases. We suggest a way of resolving the problem of the SU(3)
degeneracy of the Pauli-allowed states. A degree of the Kravchuk
polynomial serves as an additional quantum number to label the
states belonging to the same SU(3) irreducible representations.

Complete classification of the eigenfunctions and the eigenvalues
of the $^{12}$C norm kernel by the $^8$Be$=\alpha+\alpha$
eigenvalues has been given for the first time. We have
demonstrated that for the $3\alpha$ system such classification is
unique in that it is consistent with the requirements of the Pauli
exclusion principle both in the region of small intercluster
distances and in the asymptotic region where the scattering matrix
elements are produced. Due to incoincidence of eigenvalues of the
antisymmetrization operator of the $3\alpha$ system the
corresponding eigenfunctions  are uniquely determined. Any unitary
transformation applied to the latter SU(3)-basis functions would
disrupt the diagonal form of the norm kernel of $3\alpha$ system
and, hence, is inappropriate in this case. Contrastingly, the
eigenfunctions of the three-neutron or the three-boson systems are
determined only to the unitary transformation, because all the
Pauli-allowed states have unit eigenvalues. Hence, in this case
the SU(3)-classification is only one of possible variants. The
$3d'$ system is a borderline case, because the eigenvalues of the
norm kernel are equal and tend to unity for all the Pauli-allowed
states. Consequently, any transformation, which involves only the
Pauli-allowed states and remains the Pauli-forbidden states
intact, does not disrupt the diagonal form of the norm kernel.

Summarizing, the Pauli-allowed states of the $3\alpha$ system can
be arranged into the branches and the families. The eigenvalues
belonging to a given branch tend to the same limit value
$\lambda_{2k}^{{^{8}}{\rm Be}}$ of the $2\alpha$ subsystem  with
the number of quanta increasing. The branches which share limit
eigenvalues are combined in the family of the eigenstates, which
asymptotically corresponds to a certain binary decay channel of
the $3\alpha$ system into a $2\alpha$ subsystem occurring in a
particular harmonic-oscillator state and a remaining
$\alpha$-particle. Hence, the excited states of the $^{12}$C
nucleus should decay via the subsequent stage
$^{12}$C$\rightarrow^8$Be$+\alpha\rightarrow\alpha+\alpha+\alpha$
rather than in a "democratic" way
$^{12}$C$\rightarrow\alpha+\alpha+\alpha$ usually associated with
hyperspherical harmonics.

Proper truncation of the model space of the Pauli-allowed basis
functions consists in neglecting families of the states with high
values of additional quantum number $k$ and using the
angular-momentum-coupled basis in the region $\nu\gg k$ for
setting the boudary conditions for the three-cluster wave function
in the continuum.

Spectrum of the $^{12}$C norm kernel has also been compared to
that of the $^{5}$H system. The model space of the Pauli-allowed
functions in the internal region of small distance between
clusters in $3\alpha$ system has been shown to decrease
essentially as compared with the $^5$H system due to a high
symmetry of $3\alpha$-system, which comprises of three identical
clusters.

\appendix

 \section{Partial norm kernels of the $3\alpha$ system}
 \label{app:1}

The norm kernel of the $3\alpha$ system can be divided into six
partial norm kernels, each corresponding to a certain group of
nucleon permutations:
\begin{eqnarray*}
I_{3\alpha}\left({\bf A},{\bf B};\tilde{\bf A},\tilde{\bf
B}\right)=\sum_{i=1}^6{1\over3!}\hat{{\cal A}}_i\exp\left\{({\bf
A}\tilde{\bf A})+({\bf B}\tilde{\bf
B})\right\}=\sum_{i=1}^6{1\over3!}I_{3\alpha}^{(i)}\left({\bf
A},{\bf B};\tilde{\bf A},\tilde{\bf B}\right),
\end{eqnarray*}
where
\begin{eqnarray*}
I_{3\alpha}^{(1)}&=&{1\over3!}(x_1+y_1+z_1+x_2+y_2+z_2)^2,~~
I_{3\alpha}^{(2)}={1\over3!}(\bar{x}_1+\bar{y}_1+\bar{z}_1+\bar{x}_2+\bar{y}_2+\bar{z}_2)^2\\
I_{3\alpha}^{(3)}&=&{1\over3!}2(x_1+y_1+z_1+x_2+y_2+z_2)(\bar{x}_1+\bar{y}_1+\bar{z}_1+\bar{x}_2+\bar{y}_2+\bar{z}_2)
\\
I_{3\alpha}^{(4)}&=&-{1\over3!}4(x_1+y_1+z_1+x_2+y_2+z_2)(x_3+y_3+z_3)(x_4+y_4+z_4)
\\
I_{3\alpha}^{(5)}&=&-{1\over3!}4(\bar{x}_1+\bar{y}_1+\bar{z}_1+\bar{x}_2+\bar{y}_2+\bar{z}_2)
(x_3+y_3+z_3)(x_4+y_4+z_4)
\\
I_{3\alpha}^{(6)}&=&{1\over3!}4(x_3+y_3+z_3)^2(x_4+y_4+z_4)^2.
\end{eqnarray*}
We used the following notations
\begin{eqnarray*}
x_1=\exp\left\{{({\bf A}\tilde{\bf A})\over2}+{({\bf B}\tilde{\bf
B})\over2}\right\};&~~&x_2=\exp\left\{{({\bf
B}\tilde{\bf A})\over2}+{({\bf A}\tilde{\bf B})\over2}\right\};\\
y_1=\exp\left\{e^{i{2\pi\over3}}{({\bf A}\tilde{\bf
A})\over2}+e^{-i{2\pi\over3}}{({\bf B}\tilde{\bf
B})\over2}\right\},&~~&y_2=\exp\left\{e^{i{2\pi\over3}}{({\bf
B}\tilde{\bf A})\over2}+e^{-i{2\pi\over3}}{({\bf A}\tilde{\bf
B})\over2}\right\}
\\
z_1=\exp\left\{e^{-i{2\pi\over3}}{({\bf A}\tilde{\bf
A})\over2}+e^{i{2\pi\over3}}{({\bf B}\tilde{\bf
B})\over2}\right\}&~~&z_2=\exp\left\{e^{-i{2\pi\over3}}{({\bf
B}\tilde{\bf A})\over2}+e^{i{2\pi\over3}}{({\bf A}\tilde{\bf
B})\over2}\right\}\\
\bar{x}_1=2\exp\left\{-{({\bf A}\tilde{\bf A})\over4}-{({\bf
B}\tilde{\bf B})\over4}\right\};&~~& \bar{x}_2=2\exp\left\{-{({\bf
B}\tilde{\bf A})\over4}-{({\bf A}\tilde{\bf B})\over4}\right\}\\
\bar{y}_1=2\exp\left\{-e^{i{2\pi\over3}}{({\bf A}\tilde{\bf
A})\over4}-e^{-i{2\pi\over3}}{({\bf B}\tilde{\bf
B})\over4}\right\},&~~&
\bar{y}_2=2\exp\left\{-e^{i{2\pi\over3}}{({\bf B}\tilde{\bf
A})\over4}-e^{-i{2\pi\over3}}{({\bf A}\tilde{\bf
B})\over4}\right\}\\
\bar{z}_1=2\exp\left\{-e^{-i{2\pi\over3}}{({\bf A}\tilde{\bf
A})\over4}-e^{i{2\pi\over3}}{({\bf B}\tilde{\bf
B})\over4}\right\},&~~&
\bar{z}_2=2\exp\left\{-e^{-i{2\pi\over3}}{({\bf B}\tilde{\bf
A})\over4}-e^{i{2\pi\over3}}{({\bf A}\tilde{\bf
B})\over4}\right\}\\
x_3=\exp\left\{{({\bf A}\tilde{\bf A})\over4}+{({\bf B}\tilde{\bf
B})\over4}\right\};&~~&x_4=\exp\left\{{({\bf
B}\tilde{\bf A})\over4}+{({\bf A}\tilde{\bf B})\over4}\right\};\\
y_3=\exp\left\{e^{i{2\pi\over3}}{({\bf A}\tilde{\bf
A})\over4}+e^{-i{2\pi\over3}}{({\bf B}\tilde{\bf
B})\over4}\right\},&~~&y_4=\exp\left\{e^{i{2\pi\over3}}{({\bf
B}\tilde{\bf A})\over4}+e^{-i{2\pi\over3}}{({\bf A}\tilde{\bf
B})\over4}\right\}
\\
z_3=\exp\left\{e^{-i{2\pi\over3}}{({\bf A}\tilde{\bf
A})\over4}+e^{i{2\pi\over3}}{({\bf B}\tilde{\bf
B})\over4}\right\}&~~&z_4=\exp\left\{e^{-i{2\pi\over3}}{({\bf
B}\tilde{\bf A})\over4}+e^{i{2\pi\over3}}{({\bf A}\tilde{\bf
B})\over4}\right\}\\
\end{eqnarray*}

\end{document}